\documentclass[epj]{svjour}
\usepackage{graphics}
\usepackage{amsmath}
\usepackage{amstext}

\begin{document}
\title{Renormalized Parameters for Impurity Models }

\author{A.C.Hewson\inst{1} \and A. Oguri\inst{2} \and D. Meyer\inst{1}}

\institute{Department of Mathematics, Imperial College, London SW7 2BZ, UK
           \and
           Department of Material Science, Osaka City University, Sumiyoshi-ku, Osaka 558-8585 Japan
}

\date{Received: \today}

\abstract{
We show that the low energy behaviour of quite diverse impurity systems can be
described by a single renormalized Anderson model, with three parameters, an effective
level $\tilde\epsilon_d$, an effective hybridization $\tilde V$, and a quasiparticle 
interaction $\tilde U$. The renormalized parameters are calculated as a function
of the bare parameters for a number of impurity models, including
those with coupling to phonons and a Falikov-Kimball interaction
term. In the model with a coupling to phonons we determine where the 
interaction of the quasiparticles changes sign as a function of the electron-phonon coupling.
In the model with a Falikov-Kimball interaction we show that to a good approximation the
low energy behaviour 
corresponds to that of a bare Anderson model with a shifted impurity level.
}
\PACS{
{71.10.-w}
{Theories and models of many-electron systems} \and 
{71.27.+a}
{Strongly correlated electron systems; heavy fermions} \and
{75.20.Hr}
{Local moment in compounds and alloys; Kondo effect, valence fluctuations, heavy fermions}
     } 

\maketitle

\section{Introduction}

The low energy behaviour of impurities in a metallic host can be calculated using
a numerical renormalization group (NRG) approach in which the higher
energy states are progressively eliminated \cite{wil,KWW}. 
The low temperature thermodynamics of the impurity are deduced from
the leading corrections to the low energy fixed point of the renormalization
group transformation.
 An alternative approach to the calculation of the low
energy behaviour is the renormalized perturbation theory (RPT)
\cite{rpt1}.  When this approach is applied to the Anderson impurity
model, for instance, the effective low energy model generated is just
a renormalized version of the Anderson model; expressions for the
renormalized parameters can be derived in terms of the self-energy,
its derivative and the irreducible four-vertex, all evaluated at zero
frequency \cite{rpt1,rpt2}. In this paper we will clarify the relation
between these two approaches, and consider the various possible ways
of determining the renormalized parameters. We will also generalize
the approach to include models which have a coupling to phonons and a
Falikov-Kimball screening interaction between the impurity and
conduction electrons.\par
The Anderson model \cite{am} has the form,
\begin{equation} 
\begin{split}
H_{\rm AM}=&\sum\sb {\sigma}\epsilon\sb {d}
d\sp {\dagger}\sb {\sigma} d\sp {}\sb {\sigma}+ Un\sb {d,\uparrow}n\sb
{d,\downarrow}
+\sum\sb {{ k},\sigma}\epsilon\sb {{
k},\sigma}c\sp {\dagger}\sb {{ k},\sigma} c\sp {}\sb {{
k},\sigma}+\\
&+\sum\sb {{ k},\sigma}( V\sb { k}d\sp {\dagger}\sb
{\sigma} c\sp {}\sb {{ k},\sigma}+ V\sb { k}\sp *c\sp {\dagger}\sb {{
k},\sigma}d\sp {}\sb {\sigma})
.\end{split}
\label{ham}
\end{equation} 
It describes a localized level
$\epsilon_d$ of an impurity, hybridized with the conduction electrons
of the host metal via the matrix element $V_{k}$. There is in addition
a local interaction $U$ between the electrons on the impurity
site. When $U=0$ the local level broadens into a resonance,
corresponding to a localized quasi-bound state, whose width depends on
the quantity $ \Delta(\omega)=\pi\sum\sb {k}| V\sb {k}|\sp
2\delta(\omega -\epsilon\sb { k})$. It is usual to consider the case
of a wide conduction band with a flat density of states where
$\Delta(\omega)$ becomes independent of $\omega$ and can be taken as a
constant $\Delta$.\par
In the renormalized model used in the RPT approach there is an
effective energy level at $\tilde\epsilon_d$, and an effective
resonance width $\tilde\Delta$.  Expressions for these renormalized
parameters can be derived in terms of the 'bare' parameters,
$\epsilon_d$, $\Delta$, the local self-energy
$\Sigma_\sigma(\omega,h)$ and its frequency derivative
$\Sigma_\sigma'(\omega,h)$ evaluated at zero frequency $\omega=0$,
zero magnetic field $h=0$, and $T=0$. They are given by
\begin{equation}
\tilde\epsilon_{\rm d}=z(\epsilon_{\rm d}
+\Sigma_{\sigma}(0,0)),\quad\tilde\Delta =z\Delta
,\label{ren1}
\end{equation} 
where $z$, the wavefunction
renormalization factor, is given by
$z={1/{(1-\Sigma'_{\sigma}(0,0))}}$. The effective local interaction
$\tilde U$ is expressed in terms of the local four-vertex
$\Gamma_{\uparrow,\downarrow}(\omega_1,\omega_2,\omega_3,\omega_4)$
evaluated at zero frequency ($\omega_1=\omega_2=\omega_3=\omega_4=0$),
\begin{equation}\tilde U=z^2\Gamma_{\uparrow,\downarrow}(0,0,0,0).\label{ren2}\end{equation}\par
The renormalized effective model has the same form as (\ref{ham}), but
in terms of renormalized parameters, and the interaction term is
normal-ordered, as it only comes into play when two or more
excitations are created relative to the ground state of the
interacting system,
\begin{equation} 
\begin{split}
\tilde H_{\rm AM}=&\sum\sb {\sigma}\tilde\epsilon\sb {d}
d\sp {\dagger}\sb {\sigma} d\sp {}\sb {\sigma}+ \tilde U : n\sb
{d,\uparrow}n\sb {d,\downarrow}: 
+\sum\sb {{
k},\sigma}\epsilon\sb {{ k},\sigma}c\sp {\dagger}\sb {{ k},\sigma}
c\sp {}\sb {{ k},\sigma} + \\
&+\sum\sb {{ k},\sigma}(\tilde V\sb {
k}d\sp {\dagger}\sb {\sigma} c\sp {}\sb {{ k},\sigma}+\tilde V\sb {
k}\sp *c\sp {\dagger}\sb {{ k},\sigma}d\sp {}\sb {\sigma}),
\end{split}
\label{rham}
\end{equation} where the colon
brackets indicate that the expression within them must be
normal-ordered. This renormalized model is similar to that used in
earlier phenomenological local Fermi-liquid theories \cite{nh}, but
here it also includes a quasiparticle interaction term.\par
 In the renormalized perturbation theory the Hamiltonians of the bare
and renormalized Anderson models are related via $H_{\rm AM}=\tilde H_{\rm
  AM}+\tilde H_c$,
where $\tilde H_c$ is the counter-term Hamiltonian given by
\begin{equation} \tilde H_{c}=\sum\sb {\sigma}\lambda_1
d\sp {\dagger}\sb {\sigma} d\sp {}\sb {\sigma}+ \lambda_2 n\sb
{d,\uparrow}n\sb {d,\downarrow} .\label{counter}\end{equation} The
renormalized perturbation expansion is in powers of the renormalized
interaction $\tilde U$, but all the terms in (\ref{rham}) and (\ref{counter}) are taken
into account; the parameters $\lambda_1$ and $\lambda_2$, and a
rescaling factor $\lambda_3$, are determined by the condition that
there is no further renormalization of the already fully renormalized
parameters, $\tilde \epsilon_d$, $\tilde\Delta$ and $\tilde U$ arising
from the expansion (the procedure is clearer in the Lagrangian
formulation, for details see \cite{rpt2}).  It has been shown
\cite{rpt1} that the normal-ordered renormalized Anderson model
(\ref{rham}) is sufficient for the calculation of thermodynamic
properties in the low energy, low field, and low temperature regime.
The counter-term part of the Hamiltonian need only be taken into
account explicitly for the calculation of corrections for higher
temperatures and magnetic fields \cite{rpt2}. The first order
expressions for the impurity spin and charge susceptibilities,
$\chi_{s,\rm imp}$ and $\chi_{c,\rm imp}$ at $T=0$ derived from the
renormalized model are exact \cite{exact} and given by
\begin{subequations}
\begin{equation}
\chi_{s,\rm imp}={1\over 2}\tilde\rho_{\rm imp}(0)(1+
 \tilde U\tilde\rho_{\rm imp}(0)),
\end{equation}
\begin{equation}
\chi_{c,{\rm imp}}={1\over
 2}\tilde \rho_{\rm imp}(0)(1-\tilde U\tilde\rho_{\rm
 imp}(0)),
\end{equation}
\label{rsus}
\end{subequations}
where the spin susceptibility is
 given in units of $(g\mu_{\rm B})^2$ and the charge susceptibility
 differs by a factor of ${1\over 4}$ from the usual definition, so it
 is the isospin equivalent of $\chi_{s,\rm imp}$.  The quasiparticle
 density of states at the Fermi level $\tilde\rho_{\rm imp}(0)$ is
 given by
\begin{equation}\tilde\rho_{\rm imp}(0)={\tilde\Delta/\pi\over{\tilde\epsilon_d^2+\tilde\Delta^2}}\label{qpdos}.\end{equation}
 Exact results for specific heat coefficient $\gamma_{\rm imp}$ and
 the occupation of the impurity level $n_{{\rm imp},\sigma}$ are
\begin{equation}\gamma_{\rm imp}={2\pi^2\over 3}\tilde\rho_{\rm imp}(0),\label{rgam}\end{equation}
and
\begin{equation} n_{{\rm imp},\sigma}={1\over
2}-{1\over\pi}\tan ^{-1}\left ({\tilde\epsilon_{{\rm
d}}}\over{\tilde\Delta}\right ),\label{qpfsr}\end{equation} which
corresponds to the Friedel sum rule \cite{fsr}. These two results
correspond to a local Fermi-liquid theory with non-interacting
quasiparticles and can be deduced from the zero order ($\tilde U=0$)
renormalized model.  The temperature dependence of the impurity
contribution to the conductivity $\sigma(T)$ has also been calculated
for the symmetric model to order $T^2$ from the renormalized
self-energy calculated to second order in $\tilde U$. The result is
\begin{equation} \sigma(T)=\sigma_0\left\{1+{\pi^2\over
    3}\left(T\over\tilde\Delta\right)^2\left[1+2\left({\tilde
    U\over\pi\tilde\Delta}\right)^2\right]+{\rm
    O}(T^4)\right\}.\end{equation}
     When the renormalized parameters
    are expressed in terms of the self-energy and the vertex function
    these results coincide with the exact expressions derived by
    Yamada \cite{yam} from an analysis of perturbation theory to all
    orders $U$, and the Fermi-liquid results of Nozi\`eres \cite{noz}
    in the Kondo limit. This result has also been generalized to include the
leading non-linear correction to the differential conductance through a quantum dot
 in the Fermi liquid regime \cite{oguri}, which is of order $V_e^2$, where voltage $V_e$
is the voltage difference across the dot.   
However, the values of the renormalized
    parameters $\tilde\epsilon_d$, $\tilde\Delta$ and $\tilde U$, have
    to be determined. In the localized or Kondo limit they can be
    reduced to one single parameter, the Kondo temperature $T_{\rm
    K}$. For $U\gg |\epsilon_d|$, and $\epsilon_d\ll 0$, the electron
    in the d-state at the impurity site will be localized and as a
    consequence the impurity charge susceptibility must vanish in this
    limit. From equations, (\ref{rsus}) and (\ref{qpfsr}), this
    implies $\tilde U=\pi\tilde\Delta$ and $\tilde\epsilon_d=0$. If
    the impurity susceptibility at $T=0$ is expressed in terms of a
    Kondo temperature $T_{\rm K}$ defined by $\chi_{s,\rm
    imp}=1/4T_{\rm K}$, then the renormalized parameters can be
    expressed in terms of a single energy scale $T_{\rm K}$, so
    $\tilde U=\pi\tilde\Delta =4T_{\rm K}$, and the Wilson or
    $\chi/\gamma$ ratio, $R=1+\tilde U\tilde\rho_d(0)=2$
    \cite{noz}.\par
  It is clear that the renormalized Anderson model derived in the RPT
  must be directly related to the low energy effective model obtained
  in the numerical renormalization group (NRG) calculations by Wilson
  \cite{wil} for the s-d (Kondo) model, and by Krishnamurthy, Wilkins
  and Wilson (KWW) \cite{KWW} for the Anderson model.  The RPT and NRG
  approaches are, however, rather different. In the Wilson approach
  the conduction band of the Anderson model is replaced by a discrete
  spectrum of states, which is then expressed in the form of a
  tight-binding chain with the impurity at one end. The Hamiltonian
  for the discrete model for a finite chain with $N+2$ sites,
  including the impurity site, is
$$H^{N}_{\rm AM}=\sum\sb {\sigma}\epsilon\sb {d} d\sp {\dagger}\sb
{\sigma} d\sp {}\sb {\sigma}+ Un\sb {d,\uparrow}n\sb {d,\downarrow}
+V\sum\sb {\sigma}( d\sp {\dagger}\sb {\sigma} c\sp {}\sb {{
0},\sigma}+ c\sp {\dagger}\sb {{ 0},\sigma}d\sp {}\sb {\sigma})$$
\begin{equation}+\sum_{n=0,\sigma}^{n=N}\Lambda^{-n/2} \xi_n(c^{\dagger}_{n,\sigma} c^{}_{n+1,\sigma} +c^{\dagger}_{n+1,\sigma} c^{}_{n,\sigma}),\label{wham}\end{equation}
where $\Lambda >1$ is the discretization parameter, and $\xi_n$ is
given by
\begin{equation}\xi_n={D\over 2} {(1+\Lambda^{-1})(1-\Lambda^{-n-1})
\over
(1-\Lambda^{-2n-1})^{1/2}(1-\Lambda^{-2n-3})^{1/2}},\label{xi}\end{equation}
and $2D$ is the width of the conduction band \cite{wil}. The
discretization of the conduction band is logarithmic and such that the
density of levels increases as the Fermi-level is approached, and
$\epsilon=0$ is the limit point of the sequence.  With the model in
this form an iterative diagonalization scheme is then set up, starting
with the impurity site, and an extra site along the chain is added at
each iteration step. After five or six steps the matrices become too
large to handle, and after this point the Hibert space is truncated,
with the higher energy states being neglected. A fixed number, of the
order 500-1500 of the lowest lying energy states, is retained at each
subsequent step. The couplings along the chain fall off as
$\Lambda^{-n/2}$, where $n$ is the nth site as $\Lambda>1$. Lower and
lower energy scales are reached in the process but the Hamiltonian is
rescaled after each step by a factor $\Lambda^{1/2}$, so that the
lowest energy scale is formally the same after each iteration.\par
 A renormalization group transformation can be set up in which the
 states and couplings can be compared after each step. This
 renormalization group transformation for the s-d (Kondo) and Anderson
 model has a low energy fixed point corresponding to states of a free
 chain uncoupled from the impurity. In the s-d case this was
 interpreted as a $J\to\infty$ fixed point, so that the impurity and
 first site become essentially uncoupled from the chain. As this fixed
 point Hamiltonian corresponds to a free chain it gives no finite
 contribution to the impurity susceptibility and specific heat. The
 finite impurity susceptibility and specific heat arise from the
 leading irrelevant terms of the renormalization group transformation,
 which correspond to a residual interaction and a one-body coupling
 term at the end of this decoupled chain. \par
Similar results hold for the Anderson model \cite{KWW} which for the
symmetric model can be interpreted as a $V\to\infty$ fixed
point. However, from the Bethe ansatz exact solution of the Anderson
model \cite{tw,hz} we know that the Anderson model behaves as a
Fermi-liquid, whatever the value of the interaction term, so the
nature of the fixed point is independent of $U$. It seems appropriate
from a Fermi-liquid point of view, therefore, to base the
interpretation of the low energy fixed point on the original chain
including the impurity, but with renormalized parameters. In this
approach the fixed point and leading irrelevant terms are those of the
renormalized Anderson model; this then makes a direct connection with
the effective model used in the renormalized perturbation theory. It
also has the advantage that it can be interpreted in terms of a
quasiparticle picture in 1-1 comparison with the original model. We
can use the NRG technique to calculate the renormalized Fermi-liquid
parameters. A knowledge of how these renormalized parameters, such as
the quasiparticle interaction $\tilde U$, depend on the bare
parameters of the model can give us considerable insight into the low
energy behaviour of impurity systems. The numerical results for the
physical behaviour of the model at low temperatures, the
susceptibilities, specific heat coefficient etc, are essentially the same as those
obtained in the earlier NRG calculations of KWW \cite{KWW}; the
difference here is that they are interpreted in terms of a single
renormalized model, with parameters in 1-1 correspondence with those
of the original Anderson model, and the calculational procedure has also been 
considerably simplified. The approach can be extended to more
general impurity models including orbital degeneracy \cite{hew1}.\par
In the next section of this paper we give a new way of analysing the fixed point
to determine the renormalized parameters, which is straight forwardly
applicable to systems with a non-symmetric and non-constant density
of states.  The approach is then applied to
generalizations of the model to include an interaction with phonons,
and Falikov-Kimball terms, for which there are in general no exact
results. We examine how the quasiparticle interaction and renormalized
parameters vary according to the parameter regimes of these more
general impurity models. In the conclusion we discuss the question of
the extent to which these additional interactions, such as those due
to phonons and the Falikov-Kimball terms, may be absorbed by modifying
the parameters of the {\it bare model}, or whether to describe fully
the behaviour in higher energy regimes, these terms have to be
included explicitly.  \par
\section{Calculation of Renormalized Parameters}
In this section we will examine the low energy NRG fixed point of the
Anderson model as a renormalized version of the same model and deduce
the renormalized parameters $\tilde\epsilon_d$, $\tilde \Delta$ and
$\tilde U$. The starting point of the NRG calculation is the
discretized form of the model given in equation (\ref{wham}). The
many-body states are calculated using the iteration procedure as
outlined in the previous section. For a given $N$, we denote the minimum
energy required to add a single electron to  the ground
state by
$E_p(N)$, and and the minimum energy to create  a single
hole by $E_h(N)$. If these single particle and hole excitations
 correspond to a renormalized Anderson model, then
asymptotically they should coincide with those of the free model with an
appropriate choice of $\tilde\epsilon_d$ and $\tilde \Delta$ (the term in
$\tilde U$ plays a role only when more than one single particle
excitation is created from the ground state). This problem is considered in
the appendix where we define  $N$-dependent
quantities, $\tilde\epsilon_d(N)$, and $\tilde\Delta(N)=\pi\tilde V(N)^2/2$,
via the two equations,
\begin{equation}{{\pi E_p(N)\Lambda^{-(N-1)/2}}\over  2\tilde\Delta(N)}-{\pi\tilde\epsilon_d(N)\over 2\tilde\Delta(N)}=\Lambda^{(N-1)/2}g_{00}(E_p(N)),\label{rp}\end{equation}
\begin{equation}{{-\pi E_h(N)\Lambda^{-(N-1)/2}}\over 2\tilde\Delta(N)}-{\pi\tilde\epsilon_d(N)\over
   2\tilde\Delta(N)}=\Lambda^{(N-1)/2}g_{00}(-E_h(N)).\label{rh}\end{equation}
The renormalized parameters $\tilde\epsilon_d$ and $\tilde\Delta$
corresponding to the low energy fixed point are given by
$\tilde\epsilon_d=\lim_{N\to\infty}\tilde\epsilon_d(N)$ and 
$\tilde\Delta=\lim_{N\to\infty}\tilde\Delta(N)$ and are given by equations
(\ref{red}) and (\ref{rdel}) in the appendix.\par
In figure 1 we plot the quantities $\tilde\Delta(N)$ and $\tilde\epsilon_d(N)$
against $N$ for a model with $U/\pi\Delta=6.0$ and $\epsilon_d/\pi\Delta=-4.0$.
For the impurity site $N=0$ we take $\tilde\Delta(0)=\Delta$, and $\tilde\epsilon_d(0)$
to be the average Hartree-Fock value $\epsilon_d+Un_d/2$, where $n_d$
is the total occupation value at the impurity site, calculated within the
Hartree-Fock theory. This Hartree-Fock value for $\tilde\epsilon_d(0)$
is considerably shifted from the bare value $\epsilon_d$, but clearly fits the trend in 
values for $\tilde\epsilon_d(N)$ for small $N$. If the one-particle excitations  can be described by an effective non-interacting 
Anderson model then $\tilde\Delta(N)$ and $\tilde\epsilon_d(N)$
should be independent of $N$. This can indeed seen to be the case
 for $N>40$ so for this range these excitations can be described by a
model with renormalized parameters $\tilde \Delta$ and $\tilde\epsilon_d$.
As the bare parameters correspond to a model in the Kondo regime one sees that
in this case $\tilde\epsilon_d\approx 0$, whereas $\tilde\Delta/\Delta$ is small but
finite, $\tilde\Delta$  being of the order of the Kondo temperature $T_{\rm K}$.\par
Once the renormalized parameters $\tilde\epsilon_d$ and $\tilde V$ have been determined
the free quasiparticle
Hamiltonian can be diagonalized and written in the form
\begin{equation}
\Lambda^{-(N-1)/2}\sum_{k=1}^{(N+2)/2}(E_{p,k}(N)p^{\dagger}_{k,\sigma}p^{}_{k,\sigma}
+E_{h,k}(N)h^{\dagger}_{k,\sigma} h^{}_{k,\sigma}),
\label{qp}\end{equation} where $p^{\dagger}_{k,\sigma}$,
$p^{}_{k,\sigma}$, and $h^{\dagger}_{k,\sigma}$, $h^{}_{k,\sigma}$,
are the creation and annihilation operators for the quasiparticle and
quasihole excitations, and $\Lambda^{-(N-1)/2}E_{p,k}(N)$ and
$\Lambda^{-(N-1)/2}E_{h,k}(N)$ are the corresponding excitation
energies relative to the ground or vacuum state $|0\rangle$; the scale
factor $\Lambda^{-(N-1)/2}$ is due to the fact that the energies are
calculated for the rescaled Hamiltonian, which is such that
$E_{p,k}(N)$ and $E_{h,k}(N)$ for $k=1$ are of order 1. For the lowest-lying level
particle and hole levels, we have 
$E_{p}(N)=E_{p,1}(N)$ and $E_{h}(N)=E_{h,1}(N)$.\par
From the parameters, $\tilde\epsilon_d$ and $\tilde\Delta$, which determine
the free quasiparticle excitations we can immmediately deduce the occupation of the
impurity level $n_{\rm imp}$ at $T=0$ from equation (\ref{qpfsr}), which depends only on the ratio 
 $\tilde\epsilon_d/\tilde\Delta$, the quasiparticle density of states from (\ref{qpdos}), and
the impurity specific heat coefficient $\gamma_{\rm imp}$ from (\ref{rgam}).\par
To calculate the spin and charge susceptibilities and the low temperature dependence
of the conductivity we need to include a quasiparticle interaction term, 
 which for the rescaled
model takes the form,
\begin{equation} H^U(N)=
 {\tilde U}\Lambda^{(N-1)/2} :
 d_{\uparrow}^{\dagger}d_{\uparrow}^{}d_{\downarrow}^{\dagger}d_{\downarrow}^{}
 :
 \label{HU}\end{equation}
Asymptotically as $N\to\infty$ the effect of this term on the low-lying many-particle excitations
tends to zero so it is sufficient 
to calculate the effect of this term to first order in $\tilde U$. To this end we
need the operator $d_{\sigma}^{}$ expressed in terms of the
eigenstates of (\ref{qp}),
 \begin{equation}
d_{\sigma}^{}=\sum_{k=1}^{(N+2)/2}(\psi_{p,k}(-1)p_{k\sigma}+\psi_{h,k}(-1)h^{\dagger}_{k\sigma})
.\end{equation}
If the lowest two-particle excitation from the ground state for the interacting system
for a given $N$ has an energy $E_{pp}(N)$, then we can calculate $\tilde U$ by equating the
energy difference $E_{pp}(N)-2E_{p}(N)$ to that calculated using (\ref{HU}) asymptotically
in the limit $N\to\infty$. For finite $N$ we can use this equation to define an $N$-dependent
renormalized interaction $\tilde U_{pp}(N)$,
   \begin{equation}E_{pp}(N)-2E_{p}(N)=\tilde U(N)\Lambda^{(N-1)/2}
|\psi_{p,1}^*(-1)|^2|\psi_{p,1}^*(-1)|^2,
\label{tilu}\end{equation}
where $|\psi_{p,1}(-1)|^2$ is given by
\begin{equation}
 |\psi_{p,1}|^2= {1\over 1-\tilde V^2(N)\Lambda^{(N-1)}{g'}_{00}(E_p(N))}
\label{UN} 
\end{equation}
where ${g'}_{00}(\omega)$ is the derivative of $g_{00}(\omega)$.\par
Alternatively we could consider the same procedure for a two hole excitation $E_{hh}(N)$
and in a similar way define an $N$-dependent
renormalized interaction $\tilde U_{hh}(N)$, or a particle-hole excitation
 $E_{ph}(N)$ to define a
renormalized interaction $\tilde U_{ph}(N)$. In this latter case, as a positive
$U$ leads to  particle-hole attraction, we use $E_p(N)+E_h(N)-E_{ph}(N)$ on the left-hand
side of equation (\ref{tilu}).\par
If these two particle excitations can be described by an effective Anderson model then
$\tilde U_{pp}(N)$, $\tilde U_{hh}(N)$ and $\tilde U_{ph}(N)$ should be independent of
$N$ and also independent of the particle-hole labels.
In figure 1 we also plot $\tilde U_{pp}(N)$, $\tilde U_{hh}(N)$ and $\tilde U_{ph}(N)$ 
as a function of $N$ for the parameters used earlier, $U/\pi\Delta=6$ and $\epsilon_d/\pi\Delta=-4$,
with values at $N=0$ corresponding to the unrenormalized interaction $U$.
We see that for $N>40$ the values of $\tilde U_{pp}(N)$, $\tilde U_{hh}(N)$ and $\tilde U_{ph}(N)$ 
do coincide and become independent of $N$ so for the low energy excitations one can define
a unique renormalized interaction via
 $\tilde U=\lim_{N\to\infty}\tilde U_{\alpha,\alpha'}(N)$, where $\alpha,\alpha'=p,h$.
What is more, because the model with these chosen parameters corresponds to the Kondo regime, the value of $\tilde U$
coincides with the value of $\pi\tilde\Delta$, clearly seen in the inset in figure 1, and as  
can be deduced from equation (\ref{rsus}) for $\chi_{c, {\rm imp}}=0$.


\begin{figure}
\resizebox{0.95\columnwidth}{!}{%
  \includegraphics{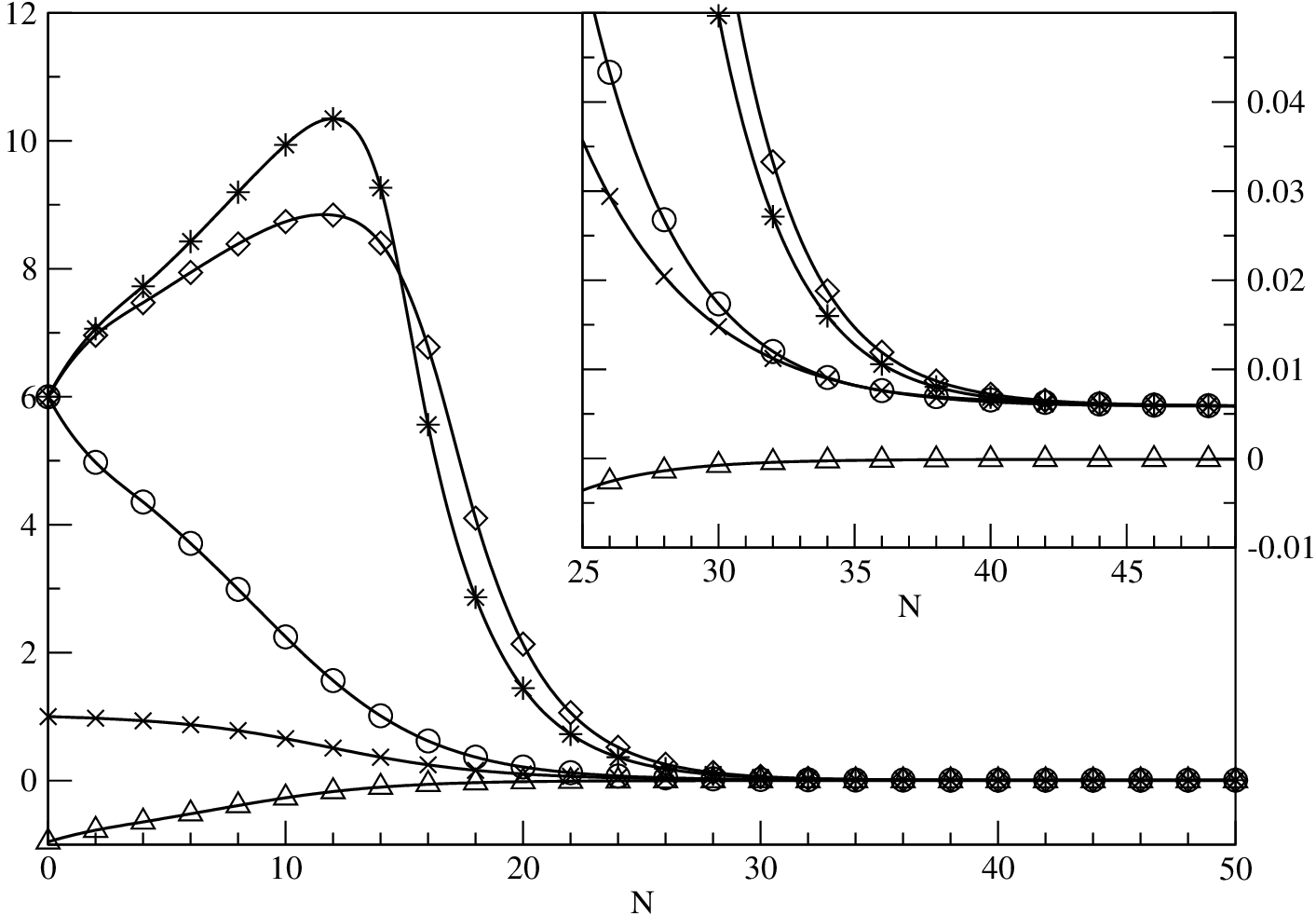}
}
\caption{Plots  of the parameters,  $\tilde\Delta(N)/\pi\Delta$ (crosses), $\tilde\epsilon_d(N)/\pi\Delta$ (triangles),
 $\tilde U_{pp}(N))/\pi\Delta$ (diamonds), $\tilde U_{hh}(N)/\pi\Delta$ (stars) and $\tilde U_{ph}(N)/\pi\Delta$ (circles), 
 for the model with bare parameters, $U/\pi\Delta=6.0$ and $\epsilon_d/\pi\Delta=-4.0$.}
\label{figure1}       
\end{figure}

\begin{table*}
\begin{center}
\begin{tabular}{|c|ccc|ccc|} \hline
NRG & & $ U/\pi\Delta=1$ & & & $U/\pi\Delta=2$ & \\ \hline & $
\tilde\Delta/\Delta$ & $\tilde U/\pi\Delta$ & R & $
\tilde\Delta/\Delta$ & $\tilde U/\pi\Delta$ & R \\ \hline
$\Lambda=2.0$ & 0.6288 & 0.4779 & 1.7600 & 0.2389 & 0.2298 & 1.9620 \\
\hline $\Lambda=2.5$ & 0.6287 & 0.4777 & 1.7599 & 0.2390 & 0.2300 &
1.9621 \\ \hline $\Lambda=3.0$ & 0.6288 & 0.4780 & 1.7601 & 0.2390 &
0.2299 & 1.9619 \\ \hline $\Lambda=3.5$ & 0.6288 & 0.4780 & 1.7601 &
0.2390 & 0.2299 & 1.9620 \\ \hline Exact (BA) & 0.6289 & 0.4780 &
1.7601 & 0.2392 & 0.2301 & 1.9620 \\ \hline
\end{tabular}
\end{center}
\caption{ The renormalized parameters for the symmetric Anderson  model for $U/\pi\Delta=1.0$ and $U/\pi\Delta=2.0$ calculated using the NRG and different values of the discretization parameter $\Lambda$.
These results are compared with the corresponding values deduced from
exact Bethe ansatz results. }
\end{table*}

\subsection{Results for Symmetric Model}

The estimates of $\tilde\Delta$ and $\tilde U$ for the symmetric model
can be checked indirectly from the exact Bethe ansatz results for this
model \cite{tw,hz}. The exact RPT results for the impurity spin
susceptibility and specific heat coefficient at $T=0$ for the
symmetric model from (\ref{rsus}) and (\ref{rgam}) are
\begin{equation}\chi_{s,\rm imp}={1\over 2\pi\tilde\Delta}\left(1+
 {\tilde U\over\pi\tilde\Delta}\right),\quad\gamma_{{\rm
 imp}}={2\pi\over 3\tilde\Delta}.\label{sym}\end{equation} By equating
 these to $\chi_{s,\rm imp}$ and $\gamma_{{\rm imp}}$ from the Bethe
 ansatz we can deduce $\tilde\Delta$ and $\tilde U$.

Though the
 discrete model (\ref{wham}) with $\Lambda>1$ and the original model
 (\ref{ham}) with a continuous spectrum have essentially the same low
 energy spectrum, i.e. they belong to the same universality class, the
 dependence of the renormalized parameters on the parameters of the
 bare model may differ. Such differences can occur in calculations
 where the high energy excitations, or high cut-offs, are treated
 differently. This situation occurs for the $N$-fold degenerate models
 ($U=\infty$) where the imposition of the high energy cut-off $D'$ in
 the Bethe ansatz calculations for the linear dispersion model differs
 from the band width $D$ for the conventional model, but a relation
 between these cut-offs can be found such that the results from the
 Bethe ansatz calculations can be translated into those for the
 conventional model \cite{rh}. A similar situation applies here.  The
 results can be made equivalent by replacing the bandwidth $D$ of the
 discrete model by $DA_{\Lambda}$, with $A_{\Lambda}$ is given by
\begin{equation}
  A_{\Lambda}={1\over 2}{1+\Lambda^{-1}\over 1-\Lambda^{-1}}{\rm
  ln}\Lambda,\label{alam}
\end{equation}
where $A_{\Lambda}\to 1$ in the continuum limit $\Lambda\to 1$.  This
result, which is given in the paper of KWW \cite{KWW} and also used in
the paper of Sakai, Shimizu and Kasuya \cite{ssk}, is derived by the
requirement that the low energy spectrum of the discrete and continuum
model coincide for $U=0$.  It
means that $V^2$ must be increased by a factor $A_{\Lambda}$ when
making a comparison with the results of a continuum model with a given
$U/\pi\Delta$.\par

\begin{figure}
\begin{center}
\resizebox{0.95\columnwidth}{!}{%
  \includegraphics{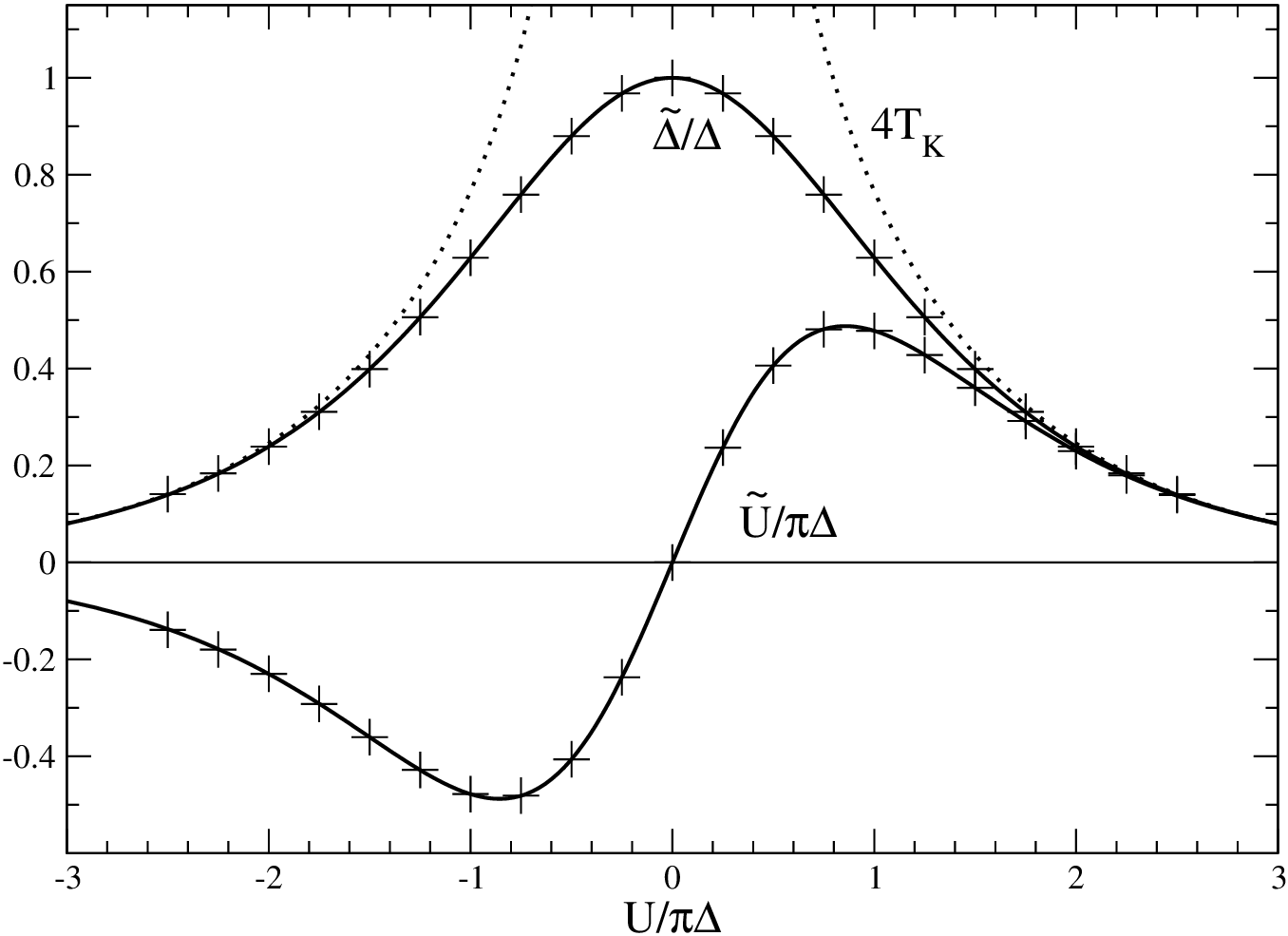}
}
\label{figure2}
\end{center}
\caption{Renormalized parameters $\tilde U/\pi\Delta$ and $\tilde\Delta/\Delta$ as a function
of $U/\pi\Delta$. The two curves are the results deduced from the
Bethe ansatz, and the points marked with a cross are those deduced
from NRG calculations with $\Lambda=2.0$. For $|U|/\pi\Delta>2.0$ the
two energy scales merge and asymptotically approach $4T_{\rm K}$
(dotted line), where $T_{\rm K}$ is the Kondo temperature.}
\end{figure}

  The values of the renormalized parameters $\tilde\Delta$,
 $\tilde U$ and the $\chi/\gamma$ or Wilson ratio, $R=1+\tilde
 U/\pi\tilde\Delta$ \cite{wil,KWW}, deduced from these are shown in
 table 1, where they are compared with those deduced from the Bethe
 ansatz results. The agreement with the Bethe ansatz results is
 remarkably good, with errors only of the order of 0.1\%, even for
 values of $\Lambda$ as large as $3.5$. It is important for such
 accurate agreement, particularly for larger values of $\Lambda$ that
 the $A_\Lambda$ factor is taken into account. \par In Figure 2 we
 make a more extensive comparison of the results for $\tilde \Delta$
 and $\tilde U$ deduced from the NRG calculations and those deduced
 from the Bethe ansatz for both positive and negative values of
 $U$. The agreement is excellent over the whole range. For
 $U/\pi\Delta>2$ the energy scales merge as the impurity charge
 becomes localized (Kondo regime) and the impurity charge
 susceptibility tends to zero. In this limit $\tilde
 U/\pi\tilde\Delta\to 1$ and as a consequence the Wilson ratio $R\to
 2$. The single renormalized energy scale in this regime is the Kondo
 temperature $T_{\rm K}$, and $\pi\tilde\Delta=\tilde U =4T_{\rm
 K}$. In the negative $U$ regime such that $U/\pi\Delta<-2$ a local
 bipolaron forms such that the spin susceptibility tends to zero, and
 the charge susceptibility is enhanced by a factor of $2$. In this
 limit the energy scales again merge such that $\tilde
 U/\pi\tilde\Delta\to -1$. This is because a Kondo effect develops in
 the isospin channel (doubly occupancy of the impurity site
 corresponding to up-isospin, and zero occupancy to down-isospin),
 such that $\pi\tilde\Delta=-\tilde U =4T_{\rm K}$, as the real spin
 fluctuations are suppressed and $\chi_{s,{\rm imp}}\to 0$. There is a
 possibility that a Kondo effect of this type might be seen in
 degenerate atomic gases with the doubly occupied paired states
 corresponding to molecules \cite{stoof}. The general expression for
 $T_{\rm K}$ that covers both Kondo regimes is
\begin{equation}T_{\rm K}=|U|\left(\Delta\over 2|U|\right)^{1/2}e^{-\pi|U|/8\Delta+\pi\Delta/2|U|}.  \end{equation}
\par
The analysis of the irrelevant terms about the fixed point is not the
only way to deduce some of the renormalized parameters from the  NRG
calculations. If the NRG approach is used also to calculate the
dynamics of the impurity model \cite{sak,chz} then $\tilde \epsilon_d$
and $\tilde \Delta$ can be deduced from the self-energy and its
derivative for $\omega=0$ and substituted in equation
(\ref{ren1}). Typical values for $\tilde \Delta$ calculated in this
way for the symmetric model ($\tilde\epsilon_d=0$) for $U/\pi\Delta=1$
and $U/\pi\Delta=2$ are 0.6155 and 0.2350, respectively for
$\Lambda=2$. They compare well with the Bethe ansatz values given in
Table 1. The errors are greater than those deduced from the analysis
of the fixed point, of the order of 2-3\%, but still quite small, and
are of the order to be expected in the calculation of dynamical
quantities.  Unfortunately the evaluation of $\tilde U$ from equation
(\ref{ren2}) requires a knowledge of the four-vertex
$\Gamma_{\uparrow,\downarrow}(0,0,0,0)$ which is very difficult to
calculate directly from an NRG calculation, as it involves the Fourier
transform of the two-particle Green's function with respect to three
independent frequency parameters. Both $\tilde\Delta$ and $\tilde U$
have, however, been calculated from equations (\ref{ren1}) and (\ref{ren2}) from perturbation
theory to third order in $U$ for the symmetric model \cite{rpt2}, and
the results are
\begin{equation} \tilde\Delta=\Delta\left\{1-\left(3-{\pi^2\over
    4}\right)\left({U\over\pi\Delta}\right)^2+\ldots\right\},
\end{equation}
and
\begin{equation} \tilde U=U\left\{1-\left({\pi^2-9  \over
    4}\right)\left({U\over\pi\Delta}\right)^2+\ldots\right\}.
\end{equation}
These perturbational results are in good agreement with the exact
results in the range $U/\pi\Delta<0.5$; at $U/\pi\Delta=0.5$ the error
in $\tilde\Delta$ is less than 1.5\%  and that for
$\tilde U$ less than 4\%.\par
We were able to define running renormalized parameters, such as
$\tilde\Delta(N)$ and $\tilde\epsilon_d(N)$ as a function of $N$, which raises 
the possibility that they can be translated into effective parameters
appropriate for calculations on an energy scale $\omega_N=\eta D\Lambda^{-(N-1)/2}$
or temperature scale, $T_N=\omega_N$, where $\eta$ is an appropriately chosen
constant of order unity. We do not however have a unique value for $\tilde U(N)$.
Nevertheless, for the particle-hole symmetric case if we take the value of
$\tilde U_{pp}(N)$ (=$\tilde U_{hh}(N)$) as $\tilde U(N)$, and translate 
this, together with $\tilde\Delta(N)$ and $\tilde\epsilon_d(N)$, into parameters
appropriate for a temperature scale $T_N$, we can generalize the RPT expression
for the impurity susceptibility in equation (\ref{rsus}) to finite temperatures, 
\begin{equation}
\chi_{\rm imp}(T)=\tilde\chi_{\rm imp}^{(0)}(T)(1+2\tilde U\tilde\chi_{\rm imp}^{(0)}(T)),\label{cht}\end{equation}
where  
$\tilde\chi_{\rm imp}^{(0)}(T)$ is the free quasiparticle contribution to the impurity
susceptibility given by
\begin{equation}
\tilde\chi_{\rm imp}^{(0)}(T)=-{1\over 2}\int_{-\infty}^{\infty}\tilde
\rho_{\rm imp}(\omega){\partial f(\omega)\over\partial\omega} d\omega\end{equation}
in units of $(g\mu_{\rm B})^2$,
where  $f(\omega)=1/(e^{\omega/T}+1)$, and $\tilde\rho_{\rm imp}(\omega)$ is the free quasiparticle
density of states given by equation (\ref{qpdos}).
 We calculate  $\tilde\chi_{\rm imp}^{(0)}(T)$  for the symmetric
model in the Kondo regime for $U/\pi\Delta=6.0$ at values of  $T_N$, using the a renormalized
parameters  $\tilde\Delta(T_N)$ and $\tilde\epsilon_d(T_N)=0$ in evaluating
the free quasiparticle density of states for this energy scale, with $\eta=1.2$ as is used in the NRG evaluation of
spectral densities on a scale $\omega_N$ (see for example \cite{sak,chz}).
We then deduce   $\chi_{\rm imp}(T)$ from equation (\ref{cht}) using
 $\tilde U(T_N)$. In figure 3 we compare the results of this calculation with the Bethe ansatz results for the s-d model given in reference \cite{tw}.
There is quite a remarkable agreement with the exact Bethe ansatz results over this temperature
range, and the value of $\chi_{\rm imp}(T)$ in the extreme high temperature range
corresponds to that of the free bare model, $1/8T$. The agreement is much less good if we
use $\tilde U_{ph}(N)$ for $\tilde U(N)$ and, as there is not a unique
prediction for this quantity, one cannot place too much reliance on this calculation. However, it does suggest that one might be able to define a renormalized perturbation theory with running coupling constants
(this is possible in a magnetic field, with field dependent parameters, as we will demonstrate elsewhere).
 \begin{figure}
\begin{center}
\resizebox{0.95\columnwidth}{!}{%
  \includegraphics{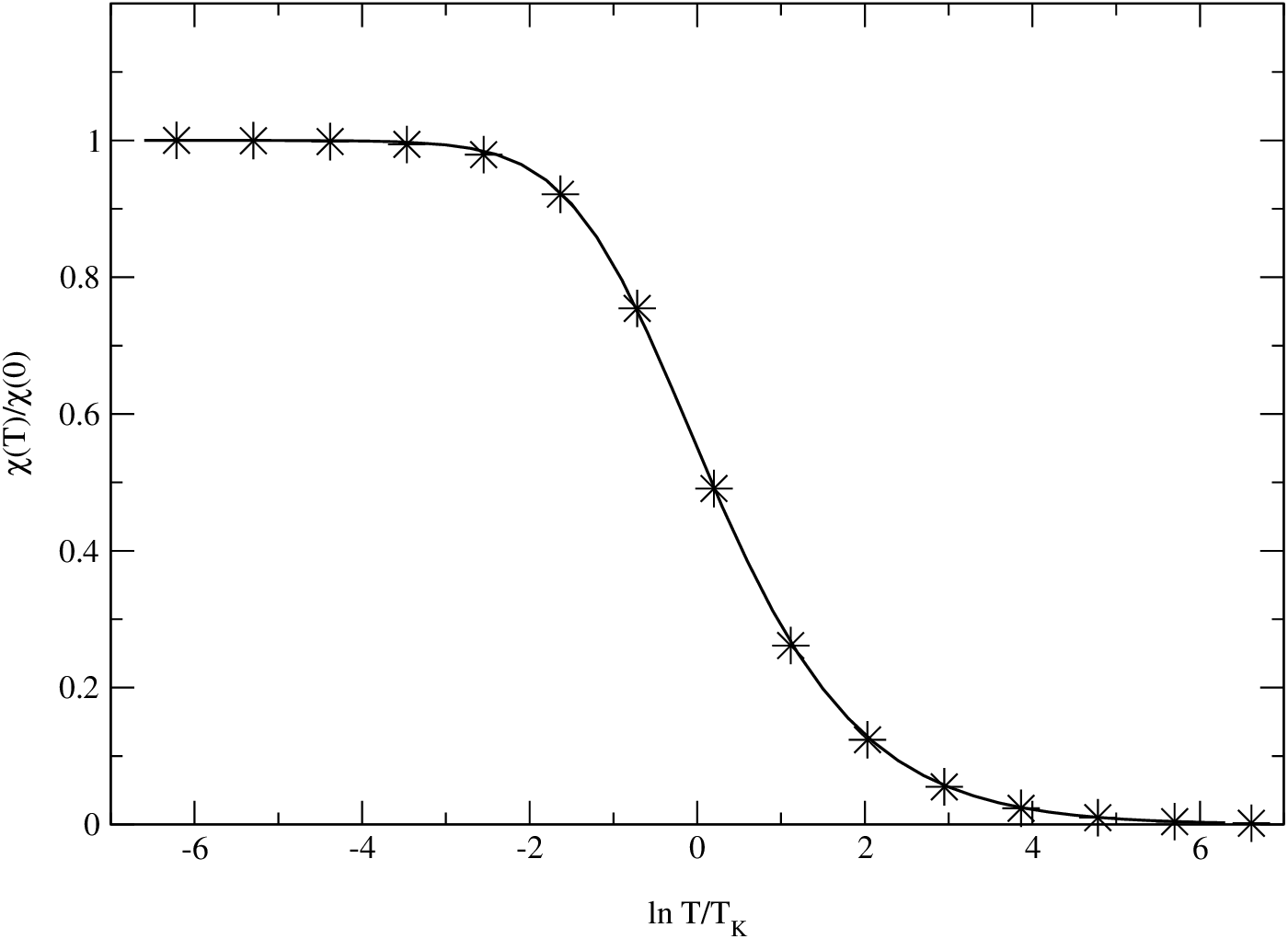}
}
\label{figurechit}
\end{center}
\caption{The temperature dependent impurity susceptibility $\chi_{\rm imp}(T)/\chi_{\rm imp}(0)$
(stars) versus ${\rm ln}(T/T_{\rm K})$ evaluated from equation (\ref{cht}) for the symmetric model  for $U/\pi\Delta=6.0$, compared to the results of the Bethe ansatz solution (continuous curve) for the s-d model given by Tsvelik and Wiegmann \cite{tw}. }
\end{figure}

\subsection{Results for Asymmetric Model}
\begin{figure}
\begin{center}
\resizebox{0.95\columnwidth}{!}{%
  \includegraphics{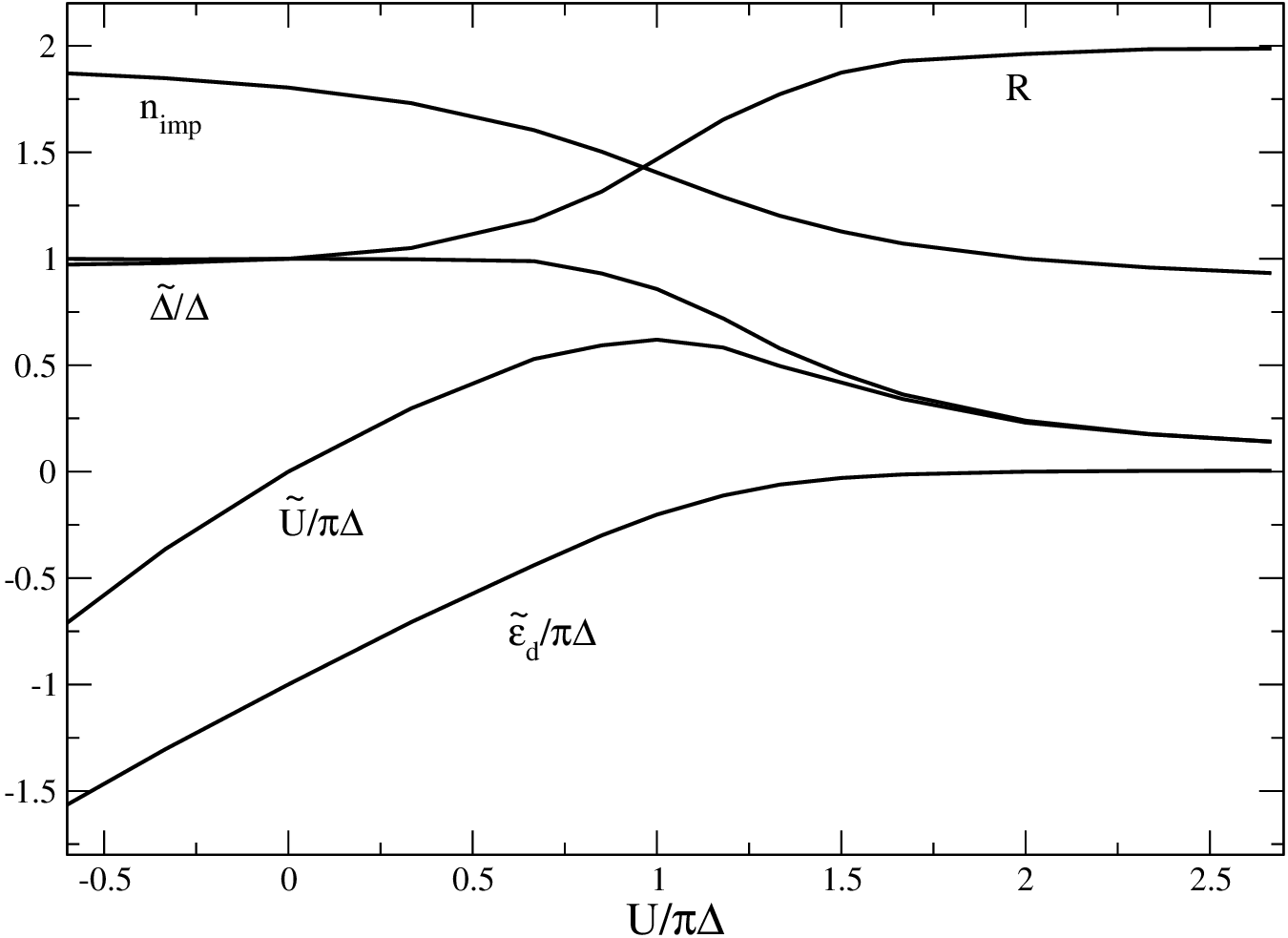}
}
\end{center}
\caption{The renormalized parameters $\tilde \Delta/\pi\Delta$, $\tilde \epsilon_d/\pi\Delta$, 
and $\tilde U/\pi\Delta$, the Wilson ratio $R=1+\tilde
U\tilde\rho_d(0)$, and the impurity occupation value $n_{\rm imp}$ are
plotted for the asymmetric Anderson model for a fixed impurity level
$\epsilon_d=-\pi\Delta$ as a function $U/\pi\Delta$. Over this range
the state of the impurity changes from the full orbital to the Kondo
regime.}
\label{figure3}
 \end{figure}
 We now consider how the renormalized parameters vary as the bare
parameters change from one qualitatively different regime of the model
to another. We start first of all with the value
$\epsilon_d=-\pi\Delta=\text{constant}$, and then increase $U$ from
$U=0$ to $U=2.7\pi\Delta$. This takes us from what might be called the
full orbital regime to the Kondo regime. The results are plotted in
figure 4. We see that $\tilde\epsilon_d$ increases at first
approximately linearly with $U$ until $U/\pi\Delta\sim 1$, and then
slowly increases monotonically through zero at the symmetric point
$U/\pi\Delta\sim 2$, remaining very close to zero in the Kondo regime
at higher values of $U$.  The renormalized resonance width
$\tilde\Delta$ decreases monotonically over the same range, though
only slowly at first in the full orbital regime, and approaches zero
in the limit $U\to\infty$. The quasiparticle interaction $\tilde U$
increases at first linearly with $U$, reaching a maximum for
$U/\pi\Delta\sim 1$, and then decreases so that its energy scale
merges with that for $\tilde\Delta$, with $\tilde
U=\pi\tilde\Delta=4T_{\rm K}$, as for the symmetric model discussed in
the previous section. Over the same parameter range $n_{\rm imp}$
decreases from an initial value of 1.8 to slightly below unity at
$U=2.7\pi\Delta$. The Wilson ratio or $\chi/\gamma$ ratio, $R=1+\tilde
U\tilde\rho_d(0)$, increases from 1 and asymptotically approaches 2 in
the Kondo regime. For negative values of $U$, $\tilde\epsilon_d$ and
$\tilde U$ decrease linearly as $U$ decreases and
$\tilde\Delta/\Delta$ approaches unity. In full orbital regime
$U/\pi\Delta<0.5$, including the range with larger negative values of
$U$, the Hartree-Fock theory, where
$\tilde\epsilon_d=\epsilon_d+Un_{d}/2$ and $\tilde\Delta=\Delta$,
constitutes a reasonably good approximation.\par It is interesting to
compare this behaviour with that for the spectral density
$\rho_d(\omega)$ of the d-electron Green's function, which can be
calculated from the NRG results. Some of the results for
$\rho_d(\omega)$ over the same parameter range are shown in figure
5(i). Initially there is only a single resonance which moves to higher
energies as $U$ is increased. As the Kondo regime is approached this
splits into a three peaked structure, the central narrow peak at the
Fermi-level being the many-body Kondo resonance. The renormalized
Anderson model given by equation (\ref{rham})
 describes only a single resonance, but is valid for the
low energy behaviour in all the parameter regimes. We see that for
smaller values of $U$, and in the negative $U$ regime,
$\tilde\epsilon_d$ tracks at first the lower resonance, increasing
monotonically, and in the Kondo regime tracks the Kondo resonance. The
precise nature of this tracking is made evident in figure 5(i) where
the curve for $\tilde\epsilon_d$ taken from figure 4 is plotted with
the spectral density results, where the main peak for each
$\rho_d(\omega)$ has been normalized to unity for the comparison; the
maxima of these peaks all lie on the curve.  The height of the
atomic-like peaks in this plot, which appear at $\omega\sim\epsilon_d$
and $\omega\sim\epsilon_d+U$ for larger $U$ in the Kondo regime,
become somewhat flattened in the normalization as the Kondo peak is so
high, making them difficult to see clearly in figure 5.\par
\begin{figure}
  \begin{center}
\resizebox{1.\columnwidth}{!}{%
  \includegraphics{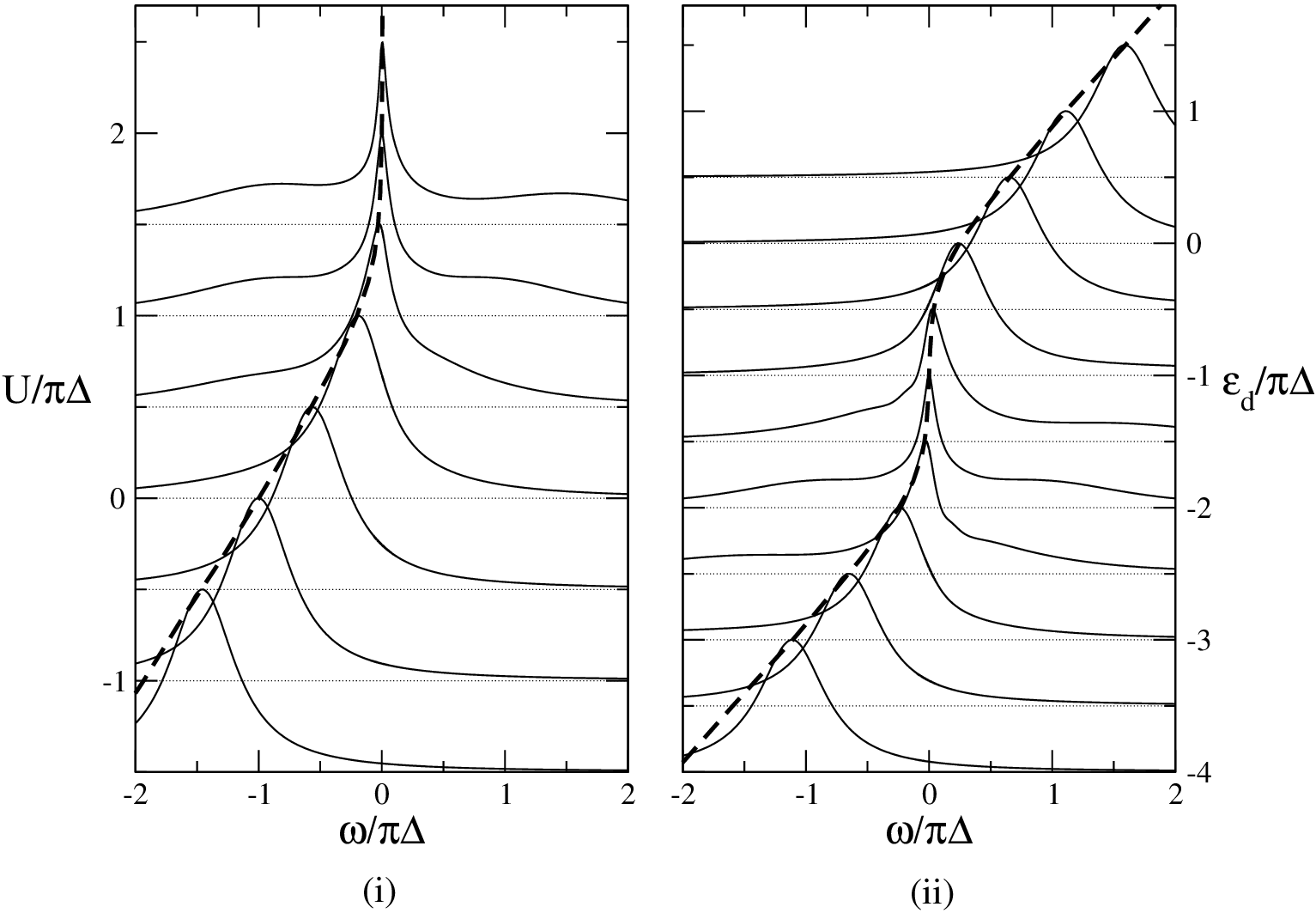}
}
  \end{center}
  \caption{ Plots of the spectral density $\rho_d(\omega)$ as a function of $\omega$
    : (i) for values of $\tilde U/\pi\Delta$ with
    $\epsilon_d=-\pi\Delta$ over the same range of values as in figure
    4, and (ii) for values of $\epsilon_d/\pi\Delta$ with $
    U=2\pi\Delta$ over the same range as in figure 6. The heights of
    the maxima in each case are normalized to unity so that the
    position of the peak can be compared with the values of
    $\tilde\epsilon_d$ (dotted line) taken from figure 4 for plot (i),
    and from figure 6 for plot (ii). The values of $U$ for the curves
    in (i) can be read off from the left hand scale from the
    coordinates of the corresponding maxima, and those for
    $\epsilon_d$ in (ii) from the right hand scale.}
  \label{figure4}
\end{figure}
 In figure 6 we give the renormalized parameters for a complementary
scan. In this case we start first of all with $\epsilon_d=-3\pi\Delta$
and $U=2\pi\Delta$, which is in the almost full orbital regime as
$\epsilon_d+U<0$, and then increase $\epsilon_d$ with a fixed value
for $U$ to a final value $\epsilon_d=1.5\pi\Delta$.  In doing so we
move to a mixed valence regime for $\epsilon_d+U\sim 0$, then for
$\epsilon_d=-\pi\Delta$ the Kondo regime for the symmetric model with
$U/\pi\Delta=2$, another mixed valence regime for $\epsilon_d\sim 0$,
and finally an empty orbital regime for $\epsilon_d>\pi\Delta$.  In
the earlier renormalization group analysis of KWW \cite{KWW}, these
parameter regimes are interpreted in terms of essentially three
different fixed points, (i) the full (empty) orbital (ii) the strong
coupling and (iii) the mixed valent. In the analysis here, however,
they are simply different parameter regimes of a single renormalized
Anderson model. We see from figure 6 the characteristic features of
full and empty orbital regimes, $\tilde\epsilon_d$ increases linearly
with $\epsilon_d$, the Wilson ratio $R\sim 1$, $n_{\rm imp}\sim 2$ or
$0$, and $\tilde\Delta$ and $\tilde U$ are independent energy scales.
In contrast in the Kondo regime, $\tilde\epsilon_d\approx 0$ and is
largely independent of $\epsilon_d$, the Wilson ratio $R\sim 2$,
$n_{\rm imp}\sim 1$, and the energy scales $\tilde\Delta$ and $\tilde
U$ have merged such that $\tilde U=\pi\tilde\Delta$.  The mixed
valence regimes appear more as cross-over regions between these two
types of behaviour, where the independence of the two energy scales
$\tilde\Delta$ and $\tilde U$ emerges.
 The curve for $\tilde\epsilon_d$ from figure 6 is plotted in figure
 5(ii) with the corresponding results for the spectral density
 $\rho_d(\omega)$, with the height of the peaks normalized to
 unity. The renormalized level $\tilde\epsilon_d$ is found to track
 the single peak in the full orbital regime as it moves up through the
 intermediate valence regime $\epsilon_d+U\sim 0$, where a second
 lower peak develops near $\omega\sim\epsilon_d$, and then it tracks
 the central narrow resonance at the Fermi-level in the Kondo regime,
 where a third upper peak develops near $\omega\sim\epsilon_d+U$. The
 reverse process occurs as $\epsilon_d$ increases from $\epsilon_d\sim
 -\pi\Delta$, as the side peaks disappear and a single peak emerges in
 the empty orbital regime. As in figure 5(i) atomic-like side peaks at
 $\omega\sim\epsilon_d$, and $\omega\sim\epsilon_d+U$, which appear in
 the mixed valence and Kondo regimes, are only just discernable due to
 the normalization of the height of the central resonance.
 If we concentrate on the mixed valence regime for $\epsilon_d\sim 0$
 in figure 6, we see that there is a significant upward shift of
 $\tilde\epsilon_d$. This reflects the effective shift in the bare
 level $\epsilon_d$ obtained by Haldane \cite{hal} in a poor man's
 scaling treatment in which the virtual charge fluctuations were
 eliminated to focus on the mixed valent regime. For the $U=\infty$
 limit, the thermodynamic behaviour of the model in this regime was
 shown by Haldane to depend on the ratio $\bar\epsilon_d/\Delta$,
 where $\bar\epsilon_d=\epsilon_d+\Delta/\pi{\rm ln}(\pi D/2\Delta)$.
 For the renormalized Anderson model the equivalent ratio is
 $\tilde\epsilon_d/\tilde\Delta$, so $\bar\epsilon_d$ is not to be
 equated with $\tilde\epsilon_d$ but $\tilde\epsilon_d/z$. In the
 mixed valence regime $\epsilon_d+U\sim 0$ the Haldane shift can be
 seen to be in the opposite direction, to lower energies.
\par
In the full and empty orbital regimes again the Hartree-Fock is a
reasonable approximation, particularly when the effective level lies
away from the Fermi-level, even though in this case we have a large
value of $U$. With $\epsilon_d=-3\pi\Delta$, $U=2\pi\Delta$,
$n_{d}=1.82$, the Hartree-Fock estimate of the position of the
renormalized level
$\tilde\epsilon_d/\pi\Delta=(\epsilon_d+Un_{d}/2)/\pi\Delta=-1.17$ is
in good agreeement with the renormalization group results, as can be
seen in figure 6. The fact that the quasiparticle interaction $\tilde
U$ is comparatively large in the full and empty orbital regimes does
not imply that the interaction effects on the low energy scale in
these regimes are large. On the contrary they are small, as can be
seen from the fact that $R$ approaches unity in these two regimes. The
reason is that the significant term is the combination $\tilde
U\tilde\rho(0)$, and in these regimes $\tilde\rho(0)\to 0$ as the
renormalized level moves well away the Fermi-level, such that $\tilde
U\tilde\rho(0)$ is quite small. A similar situation applies in the
case of the $U=\infty$ $N$-fold degenerate Anderson model where slave
boson mean field theory is asymptotically exact in the limit
$N\to\infty$. The mean field theory in that case is valid not because
$\tilde U\to 0$ as $N\to \infty$, in fact $\tilde U$ remains finite in
the limit, but because $\tilde\rho(0)\to 0$ so the combination $\tilde
U\tilde\rho(0)\to 0$ as $N\to\infty$ \cite{hew}.
\begin{figure}
  \begin{center}
\resizebox{0.95\columnwidth}{!}{%
  \includegraphics{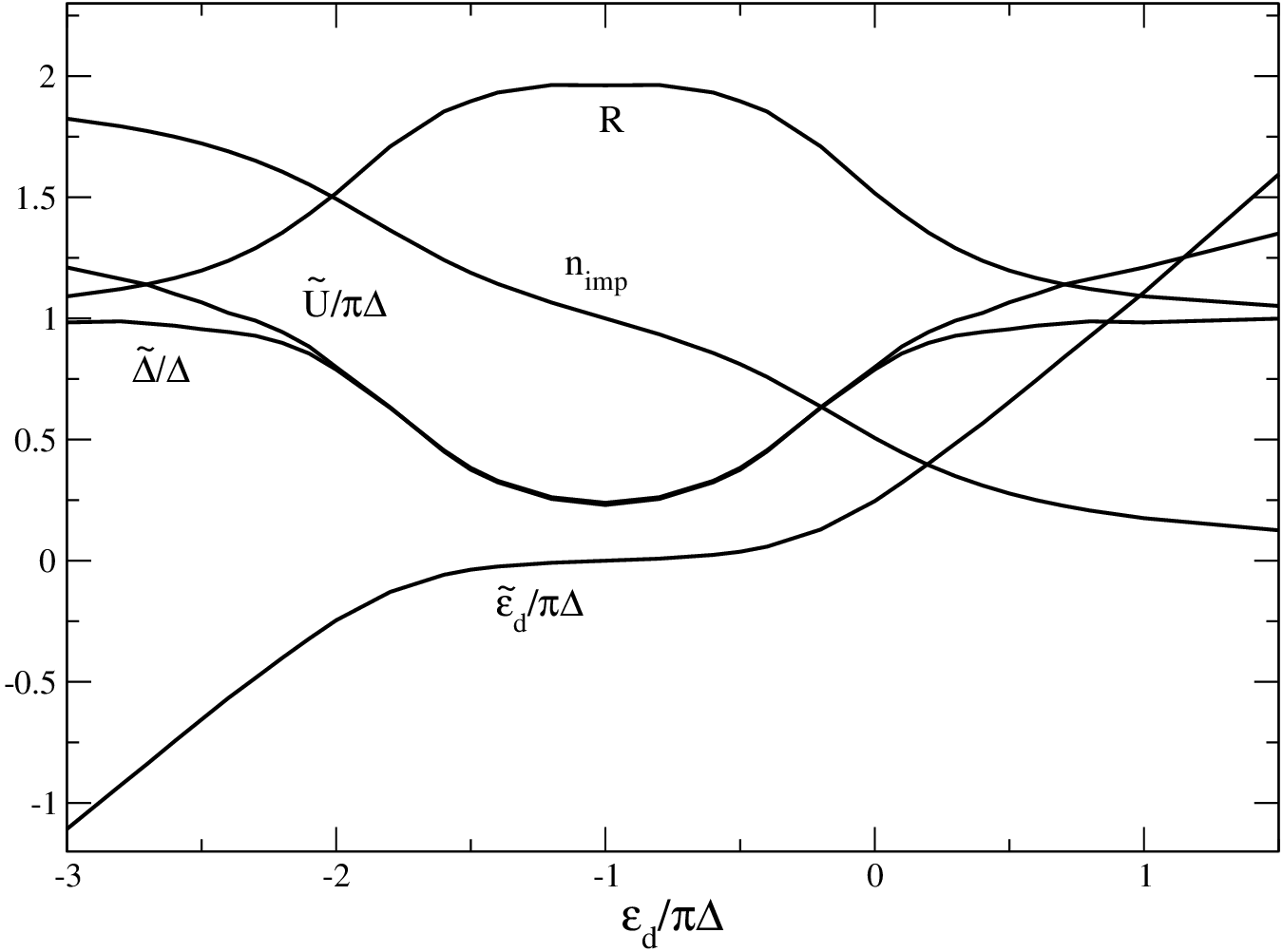}
}
  \end{center}
  \caption{ The renormalized parameters $\tilde \Delta/\pi\Delta$, $\tilde \epsilon_d/\pi\Delta$, 
    and $\tilde U/\pi\Delta$, the Wilson ratio $R$ and the impurity
    occupation value $n_{\rm imp}$ are plotted for the asymmetric
    Anderson model for a fixed value $U=2\pi\Delta$ as a function of
    $\epsilon_d/\pi\Delta$.  On increasing $\epsilon_d$ through this
    range the impurity state passes from the full orbital state,
    through a mixed valence and Kondo regime, a second mixed valence
    regime, and to a final empty orbital state.}
  \label{figure5}
\end{figure}
\par
 Though we have derived a consistent picture in the form of a
renormalized Anderson model to describe the low energy behaviour
within both the NRG and RPT approaches, the two types of calculations
involve quite different ways of realising this renormalization; it is
interesting to compare and contrast the ways in which they arrive at
the same results.  In the NRG approach the excitations at the fixed
point are free quasiparticles, and their energy spectrum depends
solely on the ratio $\tilde\epsilon_d/\tilde\Delta$, and hence via the
Friedel sum rule, equation (\ref{qpfsr}), on the additional charge at
the impurity site. The leading irrelevant terms that cause a
scattering of these quasiparticles are of the order
$\Lambda^{-(N-1)/2}$ for large $N$ \cite{com} and are of two types; a
one-body or hybridization term which determines the width
$\tilde\Delta$ of the quasiparticle resonance, and a local two-body
interaction $\tilde U$. This latter term only contributes to states in
which two or more quasiparticles are excited.\par In the RPT approach
the free quasiparticles already contain the effective hybridization,
and are characterized by the two parameters, $\tilde\epsilon_d$, the
renormalized impurity level, and $\tilde\Delta$, the renormalized
width.  As in the NRG approach the local two-body interaction term
$\tilde U$ only comes into play when two or more quasiparticles are
excited. This is because the interaction term in the quasiparticle
Hamiltonian is normal-ordered with respect to a vacuum corresponding
to the interacting ground state, in contrast to the interaction term
in the original bare model (\ref{ham}). In the more general
renormalization prescription of the RPT,
\begin{equation}
  \tilde\Sigma(0,0)=0,\quad \tilde\Sigma'(0,0)=0,\quad
  \tilde\Gamma_{\uparrow,\downarrow}(0,0,0,0)=\tilde U,
\end{equation}
where $\tilde\Sigma(\omega,T)$ is the renormalized self-energy and
the renormalized 4-vertex
$\tilde\Gamma_{\uparrow,\downarrow}(\omega_1,\omega_2,\omega_3,\omega_4)$
\cite{rpt1,rpt2}, this normal ordering is
effectively carried out by the counter terms that ensure these
conditions are satisfied.\par
The two approaches also differ in detail for analytic calculations for
the behaviour of the model at higher temperatures and higher energy
scales. In the NRG approach the next to leading order irrelevant terms
about the fixed point have to be added to the effective Hamiltonian,
and some of these will correspond to multiple quasiparticle
interaction terms.  In contrast in the RPT approach, no higher order
quasiparticle interaction terms have to be added; though the counter
terms have to be fully taken into account order by order in the
expansion in $\tilde U$. The counterpart of the multiple quasiparticle
scattering terms are the higher order Feynman diagrams in the
perturbation expansion, which have to be taken into account to probe
the higher energy scales. Calculations on these higher energy scales
should describe the 'undressing' of the fully renormalized
quasiparticles. Calculations to third order in $\tilde U$ have been
carried out for the particle-hole symmetric model
\cite{rpt2}. \par
\section{Model with a coupling to Phonons}
One possible modification of the impurity model is the inclusion of an
interaction with phonons. Such a term could be important when dealing
with an impurity with a partially filled 4f shell, as there can be a
change of the ionic volume of the order of 10\% when an electron is
removed or added to the 4f shell due to the adjustment of electrons in
the outer d and s shells. The simplest type of interaction to consider
is a coupling of the occupation of the d or f shells to an Einstein
phonon of frequency $\omega_0$, which is of the form used to study
polaronic effects in a tight-binding model by Holstein \cite{hol}. If
we add such a term to the Hamiltonian for the Anderson model we get
the Anderson-Holstein model, which we have studied in earlier work
\cite{hm}. The additional term has the form,
\begin{equation}H_{\rm e-ph}= g (b^{\dagger}+b)(\sum_{\sigma}d^{\dagger}_\sigma d^{}_\sigma -1)+ \omega_0 b^{\dagger}b,\end{equation}
where $b$ and $b^{\dagger}$ are creation and annihilation operators
for the phonon modes and a linear coupling has been assumed with a
coupling constant $g$.  Some insight into the behaviour of this model
can be obtained by performing a displaced oscillator transformation,
so that the transformed model takes the form,
\begin{equation}
\begin{split}
 H'=&\hat U^{-1}H\hat U =\sum_{\sigma} \epsilon_{d, {\rm eff}}
d_\sigma^\dagger d_\sigma^{} + U_{\rm eff}n_{d\uparrow}n_{d\downarrow}+\\
&+\sum_{{ k},\sigma}V_{\bf
  k}\left(e^{{g\over\omega_0}(b^{\dagger}-b)}d_\sigma^{\dagger} c^{}_{{ k}\sigma} +
      e^{-{g\over\omega_0}(b^{\dagger}-b)}c^{\dagger}_{{k}\sigma}d_\sigma^{}\right )+\\ 
&+\sum_{{ k}\sigma} \epsilon_{k}c^{\dagger}_{{ k}\sigma}c^{}_{{ k}\sigma} + \omega_0
      b^{\dagger}b,
\end{split}
\label{tham}
\end{equation}
where
 \begin{equation} \hat U =e^{-{g\over\omega_0}(b^{\dagger}-b)(\sum_{\sigma}n_{d,\sigma} -1)},\end{equation}
and
 \begin{equation} \epsilon_{d, {\rm eff}}=\epsilon_d+{g^2\over \omega_0},\quad U_{\rm eff}=U-{2g^2\over \omega_0}.\label{tp}\end{equation}
In the large frequency limit $\omega_0\to \infty$, such that
$g^2/\omega_0$ remains finite, the model is equivalent to the original
Anderson model with an effective level $\epsilon_{d,\rm eff}$ and an
effective local interaction $U_{\rm eff}$, because in this limit the
exponential terms that modify the hybridization term are equal to
1. When the original Coulomb interaction term $U=0$, the effective
interaction is attractive ($U_{\rm eff}<0$), so that in the localized
strong coupling regime local bipolarons form, which is equivalent to
the formation of a local moment in the spin model, and there is a
Kondo effect in the charge (or equivalently isospin) channel.\par For
the large $\omega_0$ limit, such that we can neglect the effect of the
exponential terms in the hybridization in the Hamiltonian
(\ref{tham}), we require $\omega_0$ to be much greater than the
conduction bandwidth $D$ \cite{mh}, which is unrealistic.  For the
regime $\omega_0\ll D$, the physically appropriate one, the
exponential terms in the phonon operators in (\ref{tham}) modify the
form of the original Anderson model. Nevertheless the low energy fixed
point, and the leading irrelevant interaction terms, should correspond
to a renormalized Anderson model. The explicit inclusion of the phonon
terms should only be necessary in considering higher order irrelevant
terms, which contribute only at higher energy scales and higher
temperatures. The derivation of the renormalized parameters in
equation (\ref{ren1}) holds provided that the self-energy
$\Sigma(\omega)$ is analytic at $\omega=0$. 
\begin{figure}
  \begin{center}
\resizebox{0.95\columnwidth}{!}{%
  \includegraphics{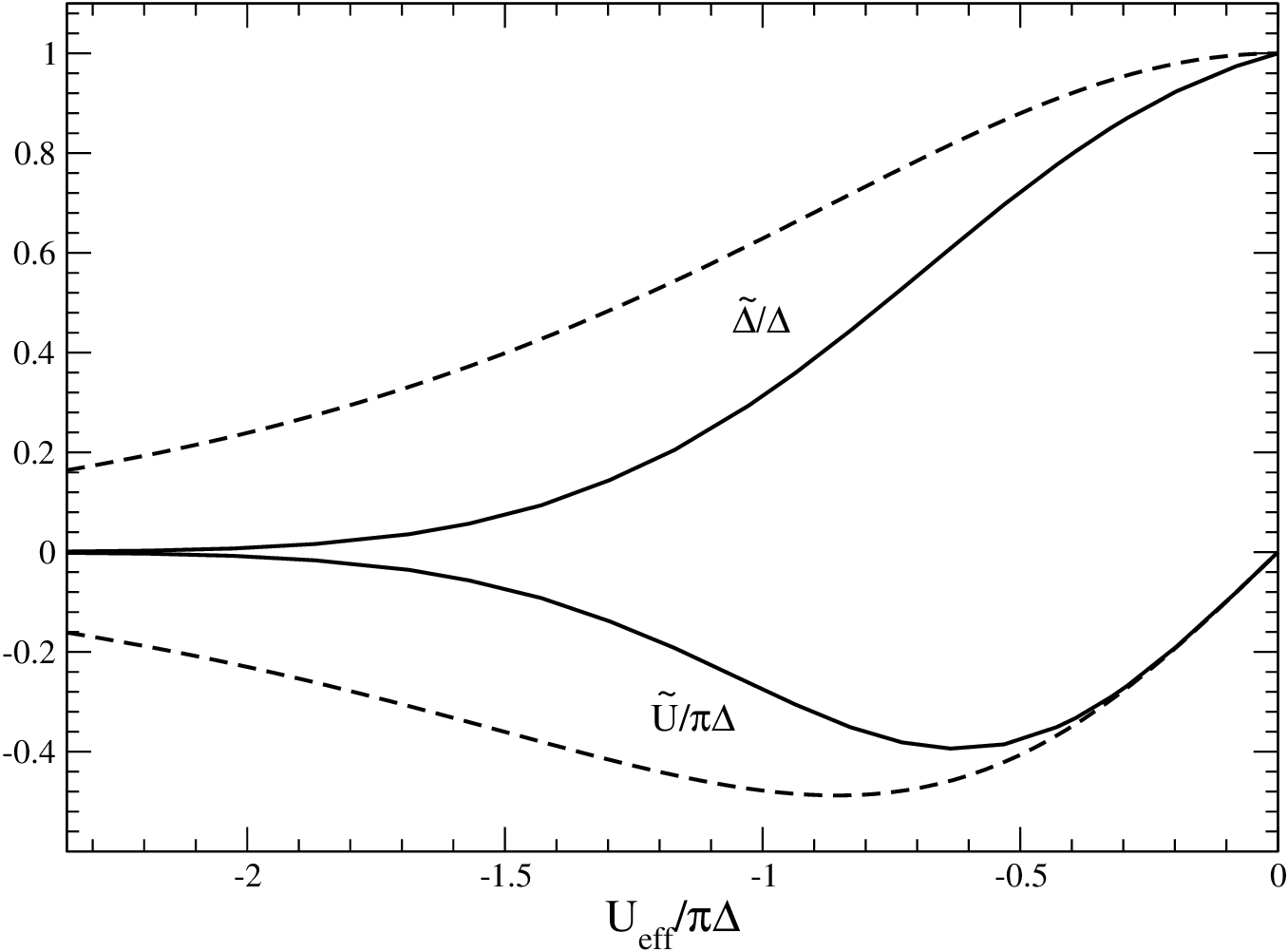}
}
  \end{center}
  \caption{ The values of the renormalized parameters $\tilde\Delta/\Delta$ and 
    $\tilde U/\Delta$ for the symmetric model with phonons ($U=0$)
    plotted as a function of $U_{\rm
    eff}/\pi\Delta=-2g^2/\omega_0\pi\Delta$. The dotted curves
    correspond to the limit $\omega_0\to\infty$ ($U_{\rm eff}$ finite)
    and are the same as those for the negative-$U$ model shown in
    figure 2. The full lines are for the corresponding curves with
    $\omega_0=0.05$.}
  \label{figure6}
\end{figure}
We can test this conjecture explicitly by again fitting the lower
energy levels obtained in the NRG calculations to a renormalized
Anderson model with parameters $\tilde\epsilon_d$, $\tilde\Delta$ and
$\tilde U$. We first of all look at the results for the particle-hole
symmetric model with $U=0$, where the phonons induce a negative-$U$
term, and assume particle-hole symmetry to compare our results with
those obtained in the previous section. In the large $\omega_0$ limit
these will be identical to those obtained earlier with $U$ being
replaced by $U_{\rm eff}=U-2g^2/\omega_0$.  For comparison we consider
the results for a value of $\omega_0\ll D$ and a commensurately
smaller value of the coupling $g$, so that it covers a similar range
of values of $U_{\rm eff}$. The results of such a fitting, using the
procedure outlined in the previous section, are shown in figure 7
for $\omega_0/D=0.05$, and for comparison also the case corresponding
to $\omega_0\to\infty$. These values
cover the negative $ U_{\rm eff}$ regime as we have $U=0$. We see
that, as the coupling $g$ increases, the values of $\tilde\Delta$
decrease much more rapidly in the smaller phonon frequency case than
in the large $\omega_0$ case due to the exponential terms that modify
the hybridization in equation (\ref{tham}). The values of $\tilde U$
are also commensurately smaller because in the localized or local
bipolaron limit, we must have $\tilde U=-\pi\tilde\Delta$, which is
the case for both sets of results in the strong coupling limit.
\begin{figure}
\begin{center}
\resizebox{0.95\columnwidth}{!}{%
  \includegraphics{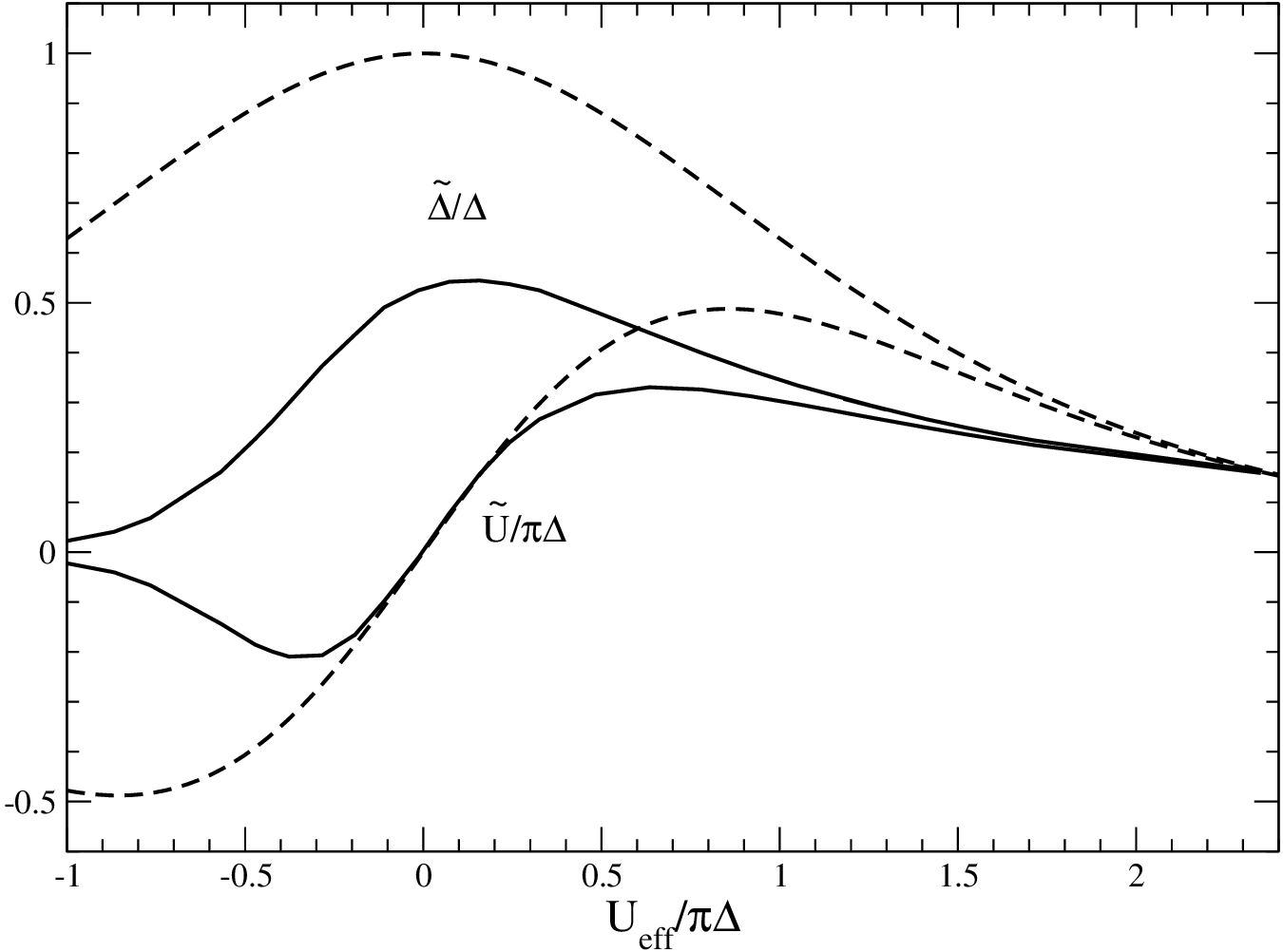}
}
\label{figure7}
 \end{center}
\caption{ The values of the renormalized parameters $\tilde\Delta/\Delta$ and 
$\tilde U/\Delta$ for the symmetric model with phonons, and
$U/\pi\Delta=2.35 $, plotted as a function of $U_{\rm
eff}/\pi\Delta=(U-2g^2/\omega_0)/\pi\Delta$. The dotted curves
correspond to the limit $\omega_0\to\infty$ ($U_{\rm eff}$ finite) and
are the same as those for the negative-$U$ model shown in figure
2. The full lines are for the corresponding curves with
$\omega_0=0.05$.}
\end{figure}
\par
In the results shown in figure 8 we start with a value of
$U/\pi\Delta=2.35$ for which the impurity charge is almost localized
and we are in the Kondo regime: as we increase $g$ one effect is to
decrease the value of $U_{\rm eff}$, and in the large $\omega_0$ limit
this is the sole effect, so that when $U_{\rm eff}=0$ we are left with
an unrenormalized width parameter $\tilde\Delta=\Delta$. In the low
$\omega_0$ case the renormalized
interaction $\tilde U$  changes sign at the same point as the
bare parameter $U_{\rm eff}$ changes sign. This result is somewhat
surprising. As the effective interaction induced by the phonons
is only valid on the scale $\omega\ll\omega_0$, one might have expected
that the on-site interaction $U$ would be significantly renormalized 
on reducing the energy scale to $\omega\sim\omega_0$, say to $\bar U$,
so that $\tilde U$ would change sign when $\bar U=2g^2/\omega_0$,
which would be significantly shifted from the condition $U_{\rm eff}=0$.
This is not the case which implies that both the direct and phonon induced
interaction terms are renormalized in a similar way.\par
 At the
point where $\tilde U=0$ there are polaronic effects, which
are absent in the large $\omega_0$ limit, such that $\tilde \Delta$ is
significantly renormalized. For the values used here the width is
reduced by factor of approximately 2 when $\tilde U=0$. In the
negative-$U_{\rm eff}$ regime the behaviour is as in figure 7, with
bipolaronic localization and $\tilde U=-\pi\tilde\Delta$, with very
much reduced values of $\tilde U$ and $\tilde\Delta$ for the small
frequency case as compared with the corresponding values for the large
$\omega_0$ limit.\par In the corresponding spinless model with a
coupling to phonons there is no induced attractive on-site
interaction, but polaronic effects occur due to the retarded effective
potential \cite{hol}. In the model with spins, we can reveal these
polaronic effects if we choose values of $U$ which cancel the
attractive interaction that leads to bipolaron formation, $U_{\rm
eff}= U-2g^2/\omega_0=0$.  We look at the renormalized resonance width
for the symmetric model for a range of values of the electron-phonon
coupling $g$ with appropriately chosen values of $U$. As a consequence
the renormalized quasiparticle interaction $\tilde U$ is negligible.
The results for the symmetric model are shown in figure 9 for
$\omega_0=0.05$ and $\omega_0=0.0125$ ($D=1$). It can be seen that the
decrease in $\tilde\Delta$ with increase of $g$ is greater in the
smaller phonon frequency case.  In the limiting case $\omega_0\to
\infty$ ($2g^2/\omega_0$ finite) there is no polaronic effect and
$\tilde\Delta/\Delta=1$ in all cases.
\begin{figure}
\begin{center}
\resizebox{0.95\columnwidth}{!}{%
  \includegraphics{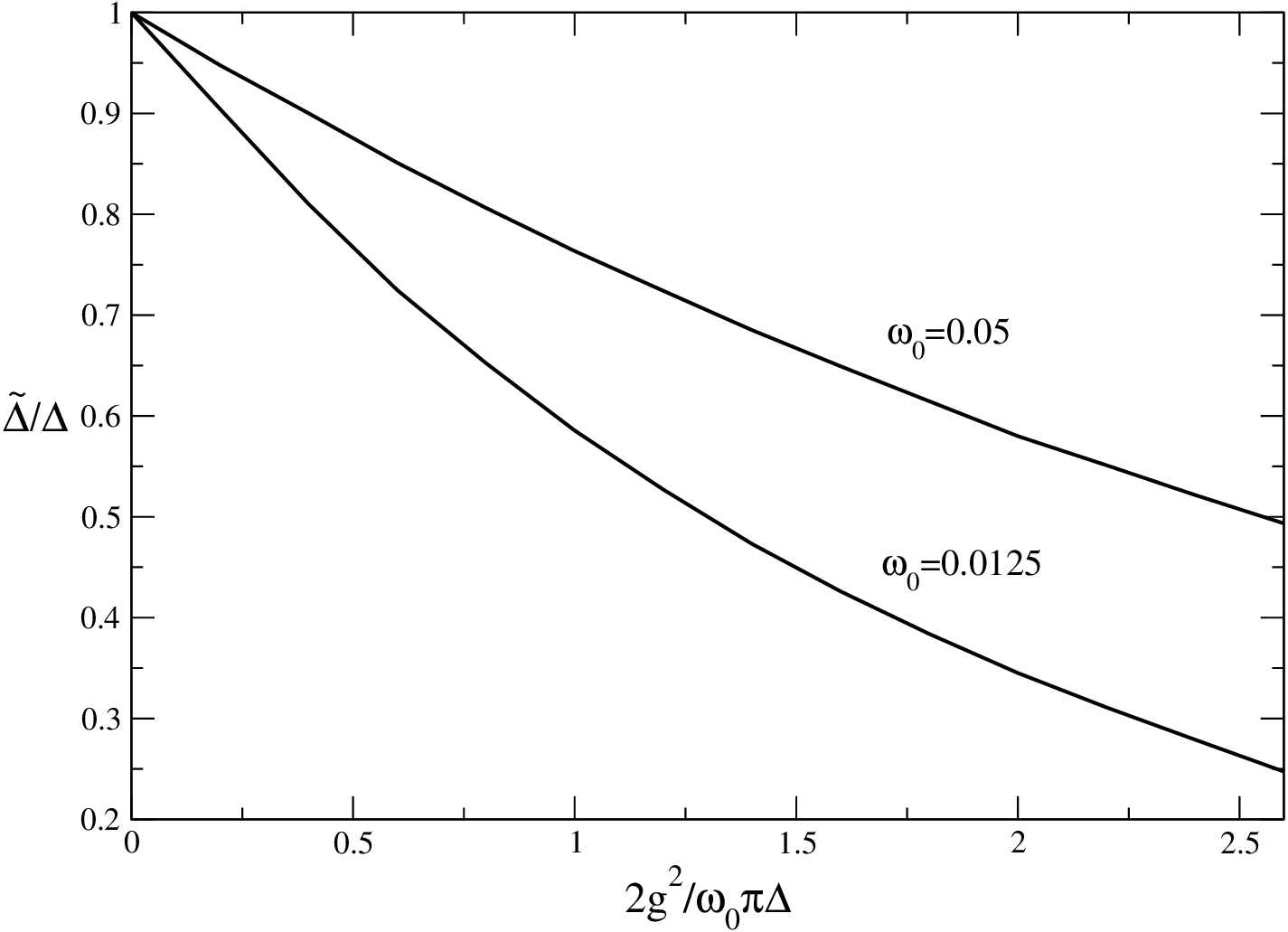}
}
\label{figure8}
\end{center}
\caption{A plot of the renormalized resonance width $\tilde\Delta/\Delta$ for the symmetric model with a phonon coupling $g$ as a function of $2g^2/\omega_0\pi\Delta$ with  values of $U$ chosen such that $U_{\rm eff}=U-2g^2/\omega_0=0$, for the two cases $\omega_0=0.05$ and  $\omega_0=0.0125$. }
   \end{figure}
Some insight into the behaviour of the asymmetric model ($U=0$) with a
coupling to phonons can be gained by comparing with the
negative-$U_{\rm eff}$ Anderson model as it corresponds to the case
$\omega_0\to\infty$. The asymmetric model with negative-$U$ can be
mapped into a positive-$U$ symmetric model with a finite magnetic
field $H\sim -\epsilon_d$ under the interchange of charge and spin;
the one-electron spectral density for the isospin up electrons is
given by $\rho_d(\omega)$ and that for the isospin down by
$\rho_d(-\omega)$.  A finite $\epsilon_d$ favours either the doubly
occupied state (isospin up) for $\epsilon_d<0$, or the empty state for
$\epsilon_d>0$, so that increasing the value of $|U_{\rm eff}|$ has
the effect of further increasing the polarization. The results for the
renormalized parameters for the asymmetric model with phonons ($U=0$
and $\epsilon_d=-\pi\Delta$) and finite $\omega_0$ ($=0.05$) which are
plotted in figure 10 show a similar trend. As $g^2$ increases, or
equivalently $U_{\rm eff}$ decreases, both $\tilde\epsilon_d$ and
$\tilde U$ decrease linearly, and the occupation of the impurity level
slowly increases and tends to the value 2. The interaction effects on
the low energy behaviour can be seen to be relatively small because,
as $\tilde\epsilon_d$ moves further from the Fermi-level, the
quasiparticle density of states at Fermi-level $\tilde\rho_d(0)$
decreases and $R\sim 1$. This trend can be clearly seen in the
calculated values of the spectral density $\rho_d(\omega)$ \cite{hm},
and $\tilde\epsilon_d$ tracks the peak in $\rho_d(\omega)$. The peaks
in $\rho_d(\omega)$ become asymmetric for strong electron-phonon
coupling, as phonon side-bands excitations are induced on the low
energy side of the peak, giving the appearance of a broadened peak.
The peak of the renormalized Anderson peak, however, is symmetric with
a width $\tilde\Delta$ slightly reduced from the bare value $\Delta$,
but is quite compatible with the spectra seen in $\rho_d(\omega)$ for
small $\omega$. The renormalized model has to reproduce the
quasiparticle spectrum only in the immediate neighbourhood of the
Fermi-level which in this case lies on the higher energy side of the
peak in $\rho_d(\omega)$.\par
\begin{figure}
\begin{center}
\resizebox{0.95\columnwidth}{!}{%
  \includegraphics{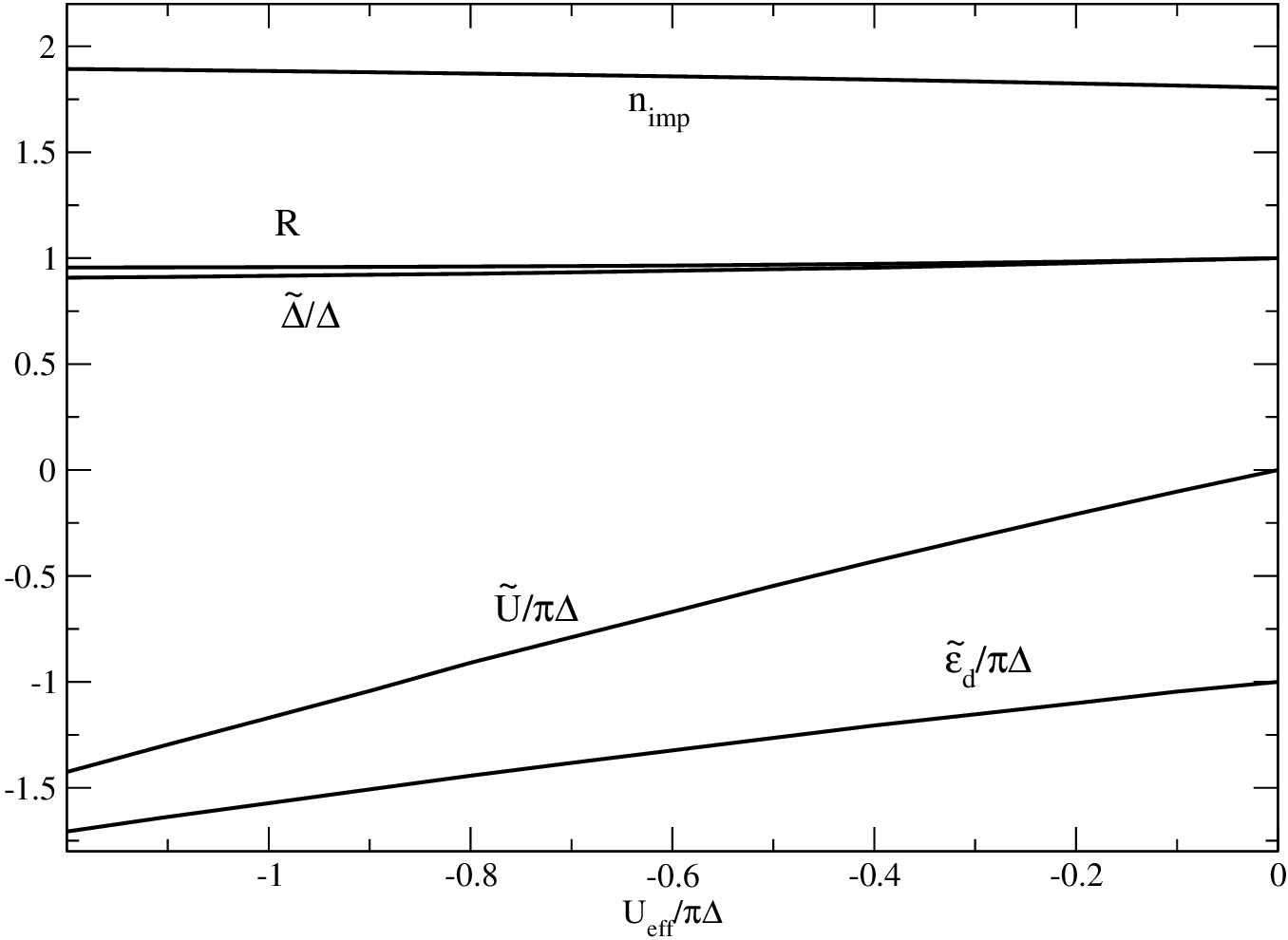}
}
\label{figure9}
\end{center}
\caption{  The renormalized parameters, $\tilde \Delta/\pi\Delta$, $\tilde \epsilon_d/\pi\Delta$, 
    and $\tilde U/\pi\Delta$, the Wilson ratio $R$ and the impurity
    occupation value $n_{\rm imp}$ plotted for the asymmetric model
    ($U=0$) with a coupling to phonons for a fixed value of
    $\epsilon_d=-\pi\Delta$ as a function of $U_{\rm eff}/\pi\Delta$
    for $\omega_0=0.05$.  }
   \end{figure}
\section{Model with a Falikov-Kimball Interaction }
Apart from the interaction $U$ between the electrons in the local d-
  or f-orbitals at the impurity site there will also be a two-body
  interaction between the electrons in these orbitals and the
  conduction electrons. Such a term does not appear in the standard
  Anderson model but a local interaction of this type is included in
  the Falikov-Kimball model \cite{fk}. It is the sole two-body
  interaction term in this model, which was put forward as an
  appropriate model for investigating valence instabilities in
  compounds, where the occupation of the localized f-orbitals changes
  significantly as a result of pressure or alloying (see reference
\cite{fz} for a recent review). The reason that
  it is not usually included in impurity models, is not because the
  interaction is very small, but because it is thought not to play an
  essential role in understanding the impurity behaviour. This means
  that its effects can, for the most part, be absorbed as a
  renormalization of terms that are already within the standard
  models. We can put this hypothesis to the test by including such a
  term in the Hamiltonian and then consider what effect it has on the
  renormalized parameters of the low energy fixed point. In the
  discrete linear chain version of the Anderson model, equation
  (\ref{wham}), used in the NRG calculations such a term can be
  included as an interaction between the impurity and the electron
  occupation at the first site on the conduction electron chain,
\begin{equation}  H_{\rm FK}=U_{\rm 
fk}\sum_{\sigma,\sigma'}n_{d,\sigma}n_{0,\sigma'},\end{equation} where
$U_{\rm fk}$ denotes the matrix element of the interaction. We add
this term to the Hamiltonian in equation (\ref{wham}) and then examine
the nature of the low energy fixed point. We take $U=2\pi\Delta$, as
earlier, with $U_{\rm fk}=U/2=\pi\Delta$, and take a range of values
of $\epsilon_d$ to run from the full orbital, through the Kondo, to
the empty orbital regime. We can then compare the results with those
obtained from a similar study in section 3.\par
\begin{figure}
  \begin{center}
\resizebox{0.95\columnwidth}{!}{%
  \includegraphics{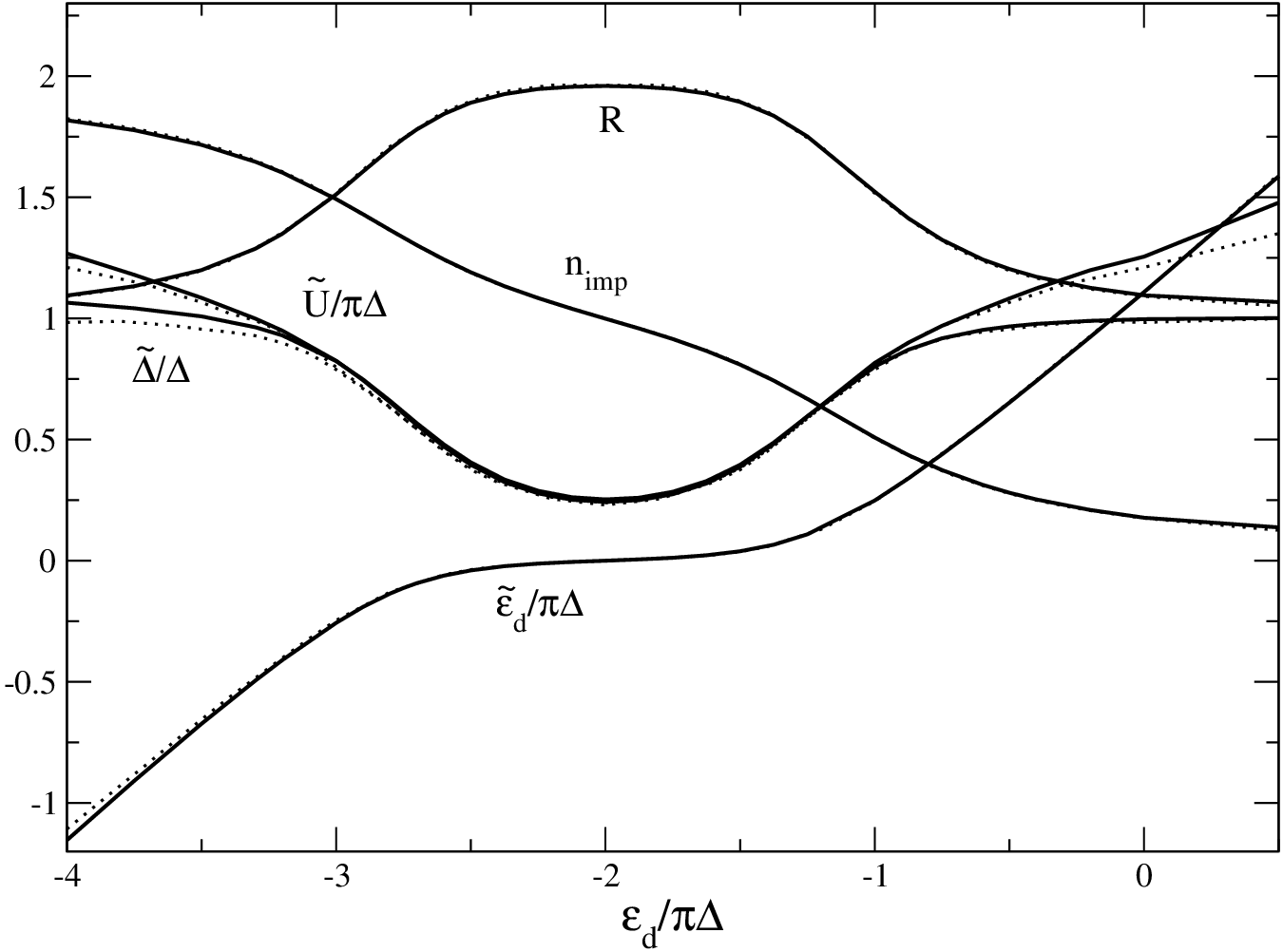}
}
  \end{center}
  \caption{ The renormalized parameters $\tilde \Delta/\pi\Delta$, $\tilde \epsilon_d/\pi\Delta$, 
    and $\tilde U/\pi\Delta$, the Wilson ratio $R$ and the impurity
    occupation value $n_{\rm imp}$ are plotted for the Anderson model
    with an additional Falikov-Kimball interaction $U_{\rm fk}$. The
    values of $U$ and $U_{\rm fk}$ are fixed, such that $U_{\rm
    fk}=U/2=\pi\Delta$, and $\epsilon_d/\pi\Delta$ is varied over the
    range indicated. Also shown as dotted curves are the renormalized
    values for the Anderson model with $U_{\rm fk}=0$, corresponding
    to those shown in figure 6, but with a shift of $\epsilon_d$ such
    that $\epsilon_d\to\epsilon_d -U_{\rm fk}$. There is a
    remarkable agreement with these displaced results over this range.
    }
  \label{figure10}
\end{figure}
  The results for the renormalized parameters in this study are shown
 in figure 11. The low temperature behaviour of the model,
$\chi_{\rm s, imp}$, $\chi_{\rm c, imp}$ and $\gamma_{\rm imp}$, can be deduced
by substituting the renormalized parameters into equations (\ref{rsus}) and
(\ref{rgam}).
There is a clear correspondence with results derived
 earlier for the Anderson model without this additional term, shown in
 figure 6. To examine the correspondence in more detail we have
 replotted the values from figure 6 in the same figure, using dotted
 lines, and have displaced the curves by a constant shift of
 $\pi\Delta=U_{\rm fk}$, such that the original bare level $\epsilon_d$ from
 figure 6 is replaced by an effective bare level
 $\bar\epsilon_d=\epsilon_d+U_{\rm fk}$.  The agreement between the
 results with the additional Falikov-Kimball term, $U_{\rm
 fk}=\pi\Delta$, and those of the model with $U_{\rm fk}=0$ and an
 effective bare level $\bar\epsilon_d$ are quite remarkable.
 In the strong correlation regime the differences are less
 than 5\%, and overall less than 10\%.  As the
 nature of the low energy fixed point has not been changed by the
 inclusion of the Falikov-Kimball term, it is not surprising that the
 low energy behaviour can still be described by a renormalized
 Anderson model.  What is somewhat unexpected in this case is that
 these renormalized parameters correspond to a constant shift of the
 bare level over the whole parameter range, and that no adjustment of
 the bare hybridization term is required. \par
To examine whether this shift varies or not with $U$, we take
$\epsilon_d=-2\pi\Delta$ with $U_{\rm fk}=\pi\Delta$ so that $\epsilon_d+U_{\rm
 fk}=-\pi\Delta$, and vary
$U$ over the same range as in the results for the pure Anderson model
shown in figure 4. If the effective bare level $\bar\epsilon_d$
is independent of $U$  the results should coincide with those
shown in figure 4.  The results of these calculations are shown in figure
12. Again we see that the results are close to
those of the Anderson model without this term, with the same shift in the
bare $\epsilon_d$. The difference in values are less than 4\% for $U/\pi\Delta>1$, and 
less than 12\% overall. In these results, and those shown in
 figure 11, the differences in the values of $n_{\rm imp}$ are even smaller, less
 than 1\%, over the whole
parameter range.\par
We conclude that the low temperature behaviour of the model with a
 Falikov-Kimball interaction $U_{\rm fk}$ in this parameter regime is very
 similar to that of an Anderson model without this term,
but with a shift of $\epsilon_d$ to $\bar\epsilon_d$. The shift is approximately independent of
 $\epsilon_d$ and $U$, and linear for $U_{\rm fk}$ in the range $0<U_{\rm
 fk}/\pi\Delta<2.0$. In this range we found the  deviations between the results of the effective
 model with $\bar\epsilon_d$ and the model with the Falikov-Kimball term
 increase  linearly with the value of $U_{\rm fk}$.
\par
\begin{figure}
  \begin{center}
\resizebox{0.95\columnwidth}{!}{%
  \includegraphics{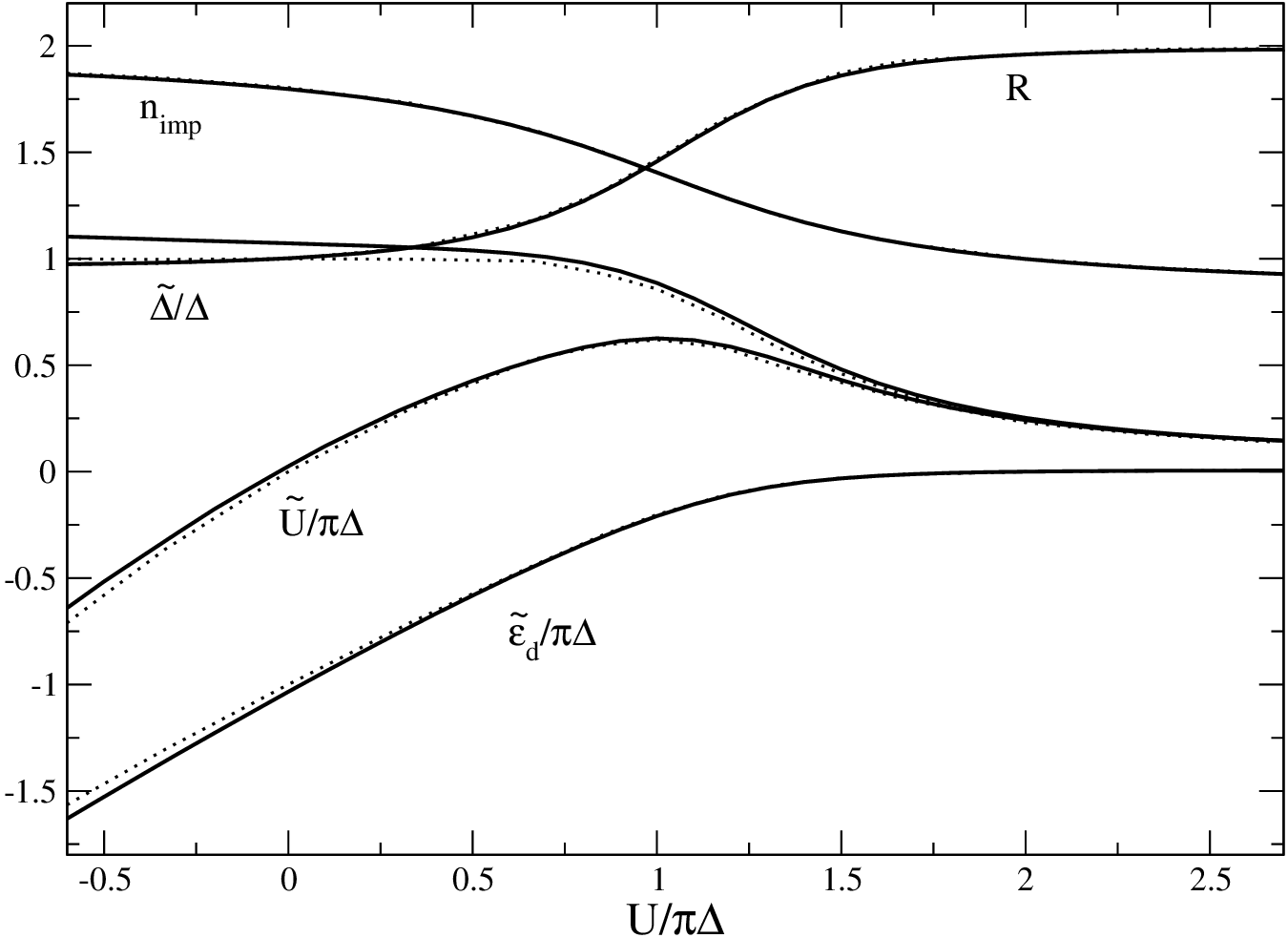}
}
  \end{center}
  \caption{ The renormalized parameters $\tilde \Delta/\pi\Delta$, $\tilde \epsilon_d/\pi\Delta$, 
    and $\tilde U/\pi\Delta$, the Wilson ratio $R$ and the impurity
    occupation value $n_{\rm imp}$ for the Anderson model
    with an additional Falikov-Kimball interaction $U_{\rm fk}$, using the same value as for figure 11. Here, $\epsilon_d=-2\pi\Delta$, and $U/\pi\Delta$ is varied over the
    range indicated. These results are compared with the corresponding values for the Anderson model with $U_{\rm fk}=0$ (dotted curves) shown in figure 4, for which $\epsilon_d=-\pi\Delta$.  There is again a
    close agreement with the results of the model without the additional term but with a displaced $\epsilon_d$.}
      \label{figure12}
\end{figure}
  In the case of a vanishing
 hybridization the nature of the fixed point Falikov-Kimball term will
 be different as it will correspond to an X-ray type of problem, with
 the localized impurity electron corresponding to a core-hole, and the
 self-energy at $\omega=0$ will be singular \cite{mahan}. This,
 however, is a special case, and in general we can expect the
 self-energy to be regular at $\omega=0$ and the renormalized Anderson
 model to provide the appropriate low energy model (see also the discussion
 in Sec. V. B of reference\cite{fz}). \par

\section{Conclusions}

We have shown that the two renormalization approaches to the
calculation of the low energy excitations of the Anderson model lead
to a consistent picture of the low energy behaviour of the model as
described by a renormalized version of the same impurity model. It is
interesting to make a comparison of this situation with field
theoretic calculations for the $\phi^4$ model or quantum
electrodynamics (QED). In the perturbation calculations for these
models infinities arise due to the lack of a high energy cut-off, but
the RPT, with the introduction of counter-terms, allows one to by-pass
these infinities and perform calculations in terms of a renormalized
version of the same model, with particle masses and interactions taken
from experiment. The RPT calculations for the Anderson model proceed
along similar lines, and the perturbation theory becomes one in terms
of the fully dressed quasiparticles and their interactions, in 1-1
correspondence with the original model. Here, however, the
renormalizations, though in some situations very large, are not
infinite as the conduction band edges provide a high energy cut-off,
and the renormalized parameters, $\tilde\epsilon_d$, $\tilde\Delta$,
the analogue of the renormalized masses can be calculated explicitly
from equation (\ref{ren1}) once the self-energy $\Sigma(\omega)$ is
known.\par
The Anderson model, as a model for impurity systems,
neglects many higher energy scale interactions. In including an
electron-phonon coupling and a Falikov-Kimball term we have included
some of the possible higher energy interaction terms. Explicit
calculations have shown that these terms do not change any essential
features of the low energy behaviour, which can still be described in
terms of a three-parameter renormalized Anderson model, though the
renormalized parameters are modified.  It is reasonable to assume that
if other types of interaction terms are included, such as two-body
hybridization terms arising from off-diagonal elements of the Coulomb
interaction of the impurity d-electron with the conduction electrons,
that the same would hold.  The fact that we can describe the various
forms of low energy behaviour of this class of non-degenerate impurity
models, within the framework of a single renormalized model, gives a
unifying perspective. It also simplifies the interpretation of experiments
on the low temperature behaviour of impurity systems if the results
can be analysed in terms of a single three-parameter model.\par
 The non-degenerate Anderson model is not
sufficient to describe many impurity systems where the impurity ion
has localized incomplete d or f-shells. There are generalizations of
the Anderson model that take account of the degeneracy of these
configurations, such as the $U=\infty$ N-fold degenerate model for 4f
ionic systems, and the n-fold denenerate models with Hund's rule
coupling for d-state ions, and renormalized versions of these models
also describe the low energy regime.  Explicit calculations of the
renormalized parameters have only been carried out in the Kondo
regime, with renormalized parameters, such as the renormalized Hund's
rule coupling expressed in terms of $T_{\rm K}$
\cite{hew1,hew}. Models that have a non-Fermi-liquid low energy fixed
point, however, are outside the scope of the present discussion.\par
Though the extra interactions, such as the coupling to phonons or the
Falikov-Kimball term, need not be included explicitly to describe
the very low energy behaviour, the question arises as to whether they
are  needed to describe the
behaviour on higher energy scales, or whether can they be simply taken into
account via a re-parameterization of the bare model to
$\bar\epsilon_d$, $\bar\Delta$ and $\bar U$ \cite{hew,hbby}.  There is no simple
answer to this question. There are situations where this is formally
possible, such as in the model with coupling to phonons in the
$\omega_0\to\infty$ limit, where the phonons can be effectively
eliminated to give a renormalized bare model, as given in equation
(\ref{tp}). However, for $\omega>\omega_0$ the real phonon excitations,
and not just virtual ones, have to be taken into account explicitly.
 Even the interaction term $U$ of the bare Anderson model
is a renormalized one because its experimental value
does not correspond to the Coulomb matrix elements for electrons in
localized d-orbitals, but takes into account the many-body
relaxation and screening effects of the other electronic shells, when
a d or f-electron is removed; so the model is already a
renormalized or effective model. This is true for most bare models: which
interactions have to be taken into account explicitly is dependent on
the energy scales and the type of experiment which is being
described.\par

\begin{acknowledgement}
One of us (ACH) wishes to thank the EPSRC (Grant GR/S18571/01) for
financial support, and W. Koller for helpful advice. AO wishes to acknowledge 
 support by the 
Grant-in-Aid
for Scientific Research from JSPS.  We are also
grateful to Y. Shimizu and O. Sakai for helpful discussions, and
particularly in relation to the derivation of the correction factor
$A_{\Lambda}$ for the discretized model.
\end{acknowledgement}

\appendix

\section{Appendix}

Let the linear chain Hamiltonian (\ref{wham}) of  the non-interacting
system plus impurity be denoted by $H^0_{-1,N}$, and that of the rest of the chain without
the impurity starting at site $i$ by $H^0_{i,N}$, where $i=0,1,2\ldots $. We are considering
free single particle excitations  for $\tilde U=0$
so that the Green's function at the impurity site ($i=-1$) is given by
\begin{equation}G_{-1-1}(\omega)={1\over \omega-\epsilon_d\Lambda^{(N-1)/2}- V^2\Lambda^{(N-1)}
g_{0 0}(\omega)}\label{gimp}\end{equation}
where $g_{0 0}(\omega)$ is the Green's function at site $i=0$ for the system described by the Hamiltonian $H^0_{0,N}$. This Green's function is given in turn by
 \begin{equation}g_{00}(\omega)={1\over \omega-\xi_0^2\Lambda^{(N-1)}g_{1 1}(\omega)}\label{eq1}\end{equation}
and $g_{1 1}(\omega)$ is the Green's function at site $i=1$ for the system
described by the Hamiltonian $H^0_{1,N}$. This procedure can be extended to express $g_{00}(E)$ in the form of a continued fraction. \par 
The one-particle excitations  $E$ are given by the poles of (\ref{gimp}),
\begin{equation}E-\epsilon_d\Lambda^{(N-1)/2}- V^2\Lambda^{(N-1)}g_{00}(E)=0,\label{poles}\end{equation}
where  $g_{00}(E)$ can be expressed in the form,
\begin{equation}g_{00}(E)=\sum_{{l}=1,N+1}{ |\phi_{l}(0)|^2\over E-\tilde E_{l}(N)},\label{g00}\end{equation}
where $\phi_{l}(0)$ and $\tilde E_{l}(N)$ are the eigenvectors and eigenvalues
of $H^0_{0,N}$.  
\par
If $E^0_p(N)$ and $E^0_h(N)$ are the lowest particle and hole excitations  from the
ground state of the Hamiltonian $H^0_{-1,N}$ then,
\begin{equation}{{E^0_p(N)\Lambda^{-(N-1)/2}}\over  V^2}-{\epsilon_d\over  V^2}=\Lambda^{(N-1)/2}g_{00}(E^0_p(N)),\label{0p}\end{equation}
\begin{equation}{{-E^0_h(N)\Lambda^{-(N-1)/2}}\over  V^2}-{\epsilon_d\over
    V^2}=\Lambda^{(N-1)/2}g_{00}(-E^0_h(N)).\label{0h}\end{equation}
If we replace $E^0_{p(h)}$ by the corresponding values for the interacting
system $U\ne 0$,  $E_{p(h)}$, then they can be used to define  $N$-dependent
quantities, $\tilde\epsilon_d(N)$, and $\tilde\Delta(N)=\pi\tilde V(N)^2/2$,
\begin{equation}{{\pi E_p(N)\Lambda^{-(N-1)/2}}\over  2\tilde\Delta(N)}-{\pi\tilde\epsilon_d(N)\over 2\tilde\Delta(N)}=\Lambda^{(N-1)/2}g_{00}(E_p(N)),\label{polesp}\end{equation}
\begin{equation}{{-\pi E_h(N)\Lambda^{-(N-1)/2}}\over 2\tilde\Delta(N)}-{\pi\tilde\epsilon_d(N)\over
   2\tilde\Delta(N)}=\Lambda^{(N-1)/2}g_{00}(-E_h(N)).\label{polesh}\end{equation}
The renormalized parameters $\tilde\epsilon_d$ and $\tilde\Delta$
corresponding to the low energy fixed point are given by
$\tilde\epsilon_d=\lim_{N\to\infty}\tilde\epsilon_d(N)$ and 
$\tilde\Delta=\lim_{N\to\infty}\tilde\Delta(N)$, and are determined by the
two equations,
\begin{equation} {\pi\tilde \epsilon_d\over{2\tilde
      \Delta}}=\lim_{N\to\infty}\Lambda^{(N-1)/2}g_{00}(\pm E_{p(h)}(N)),\label{red}\end{equation}
and
\begin{equation}{\pi\over 2\tilde\Delta}=\lim_{N\to\infty}{\Lambda^{(N-1)}(g_{00}(E_p(N))-
g_{00}(-E_h(N)))\over E_p(N)+E_h(N)}.\label{rdel}\end{equation}\par
The results in equations (\ref{red}) and (\ref{rdel}) are applicable also  systems with
non-symmetric or non-constant densities of conduction states, with the appropriate form 
for $g_{00}(\omega)$, which can be evaluated using either equation (\ref{g00}) or
(\ref{eq1}).




\begin{thebibliography}{}
\bibitem{exact} That these results are exact can be deduced from a
Ward identity. For details see references \cite{yam,rpt1}.\par
\bibitem{com} This is a feature of the discrete model, more generally
in a continuum case one might expect these terms to fall off as
$1/N$.\par
\end{thebibliography}
\end{document}